\documentclass[
preprint,
showpacs,
 amsmath,amssymb,
 aps,
 prd,
 superscriptaddress,
]{revtex4-1}

\usepackage{graphicx}
\usepackage{dcolumn}
\usepackage{bm}

\usepackage{color}
\usepackage{hyperref}
\usepackage{caption}
\usepackage{subcaption}
\usepackage{url}

\usepackage{booktabs} 
\usepackage{array} 
\usepackage{paralist} 
\usepackage{verbatim} 

\usepackage{fancyhdr} 
\usepackage{amsmath}
\usepackage{graphicx}
\usepackage{amsthm}
 
\theoremstyle{definition}

\DeclareMathOperator{\Tr}{Tr}

\DeclareMathOperator{\co}{:}

\newcommand{\be}{\begin{equation}}
\newcommand{\ee}{\end{equation}}
\newcommand{\bea}{\begin{eqnarray}}
\newcommand{\eea}{\end{eqnarray}}
\newcommand{\area}{\mathcal{A}}
\newcommand{\tG}{\widetilde{G}}
\newcommand{\overbar}[1]{\mkern 1.5mu\overline{\mkern-1.5mu#1\mkern-1.5mu}\mkern 1.5mu}

\begin{document}

\preprint{CALT 2016-15}

\title{Space from Hilbert Space: \\  Recovering Geometry from Bulk Entanglement}


\author{ChunJun \surname{Cao}}
\email{cjcao@caltech.edu}
\affiliation{Walter Burke Institute for Theoretical Physics California Institute of Technology, Pasadena, CA 91125, USA}
 
\author{Sean M. \surname{Carroll}}
\email{seancarroll@gmail.com}
\affiliation{Walter Burke Institute for Theoretical Physics California Institute of Technology, Pasadena, CA 91125, USA}

\author{Spyridon \surname{Michalakis}}
\email{spiros@caltech.edu}
\affiliation{Walter Burke Institute for Theoretical Physics California Institute of Technology, Pasadena, CA 91125, USA}
\affiliation{Institute for Quantum Information and Matter, California Institute of Technology, Pasadena, CA 91125, USA}


\begin{abstract}
We examine how to construct a spatial manifold and its geometry from the entanglement structure of an abstract quantum state in Hilbert space. 
Given a decomposition of Hilbert space $\mathcal{H}$ into a tensor product of factors,  we consider a class of ``redundancy-constrained states" in $\mathcal{H}$ that generalize the area-law behavior for entanglement entropy usually found in condensed-matter systems with gapped local Hamiltonians. Using mutual information to define a distance measure on the graph, we employ classical multidimensional scaling to extract the best-fit spatial dimensionality of the emergent geometry. 
We then show that entanglement perturbations on such emergent geometries naturally give rise to local modifications of spatial curvature which obey a (spatial) analog of Einstein's equation. 
The Hilbert space corresponding to a region of flat space is finite-dimensional and scales as the volume, though the entropy (and the maximum change thereof) scales like the area of the boundary.
A version of the ER=EPR conjecture is recovered, in that perturbations that entangle distant parts of the emergent geometry generate a configuration that may be considered as a highly quantum wormhole.
\end{abstract}

\maketitle

\baselineskip=15pt

\tableofcontents
\newpage

\section{Introduction}

Quantum-mechanical theories are generally thought of as theories \emph{of} something.
Quantum states are square-integrable complex-valued functions of the configuration of some particular kind of ``stuff,'' where that stuff may be a simple harmonic oscillator, a set of interacting spins, or a collection of relativistic fields.

But quantum states live in Hilbert space, a complete complex vector space of specified dimension with an inner product.
The same quantum states, even with the same dynamics, might be thought of as describing very different kinds of stuff.
Coleman long ago showed that the quantum theory of the sine-Gordon boson in 1+1 dimensions was equivalent to that of a massive Thirring fermion \cite{PhysRevD.11.2088}.
AdS/CFT posits an equivalence (in a certain limit) between a conformal field theory in a fixed $d$-dimensional Minkowski background and a gravitational theory in a dynamical $(d+1)$-dimensional spacetime with asymptotically anti-de~Sitter boundary conditions \cite{Maldacena:1997re}.
The wave functions of a single quantum theory can be represented in very different-looking ways.
It is therefore interesting to consider the inverse problem to ``quantizing'' a theory: starting with a quantum theory defined in Hilbert space, and asking what it is a theory of. 
In this paper we take steps toward deriving the existence and properties of space itself from an intrinsically quantum description using entanglement.

A good deal of recent work has addressed the relationship between quantum entanglement and spacetime geometry.
Much of the attention has focused on holographic models, especially in an AdS/CFT context.
Entanglement in the boundary theory has been directly related to bulk geometry, including deriving the bulk Einstein equation from the entanglement first law (EFL) 
 \cite{relativeentropyandholography,graventanglement1,graventanglement2,graventanglement3}. 
(The EFL relates a perturbative change in the entropy of a density matrix to the change in the expectation value of its modular Hamiltonian, as discussed below.)
Tensor networks have provided a connection between emergent geometry, quantum information, and many-body systems \cite{tensorandgeometry,Swingle:2009,Swingle:2012,qi,Bao:2015, Czechetal:2015oct,Czechetal:2015dec,HaPPY:2015,Hayden:2016}.

It is also possible to investigate the entanglement/geometry connection directly in a spacetime bulk.
The ER=EPR conjecture relates entanglement between individual particles to spacetime wormholes  \cite{er-eprms,er-eprmvr,er-eprfluxtube,Baoetal:2015jul,Baoetal:2015sept,Baoetal:2016apr}.
Consider two entangled particles, separated by a long distance, compared to the same particles but unentangled. 
If sufficient entanglement gives rise to a wormhole geometry, some weak gravitational effects should arise from small amounts of entanglement, and evidence for this phenomenon can be found in the context of AdS/CFT  \cite{Bruschi,er-eprfluxtube,er-eprmvr}. 
From a different perspective, Jacobson has argued that Einstein's equation can be derived from bulk entanglement under an assumption of local thermodynamic equilibrium between infrared and ultraviolet degrees of freedom \cite{jacobson15,Casini:2016rwj,Carroll:2016lku}.

While the current paper is inspired by the idea of emerging space from entanglement, our approach of bulk emergent gravity differs from the aforementioned papers in that our starting point is directly in Hilbert space, rather than perturbations around a boundary theory or a semiclassical  spacetime.
We will first try to construct a generic framework by which an approximate sense of geometry can be defined purely from the entanglement structure of some special states. We conjecture that mutual information  (See \cite{preskillLecture,Nielsen:2011:QCQ:1972505} for a review), similar to suggestions by \cite{er-eprmvr,qi}, can be used to associate spatial manifolds with certain kinds of quantum states.
More tentatively, we explore the possibility that perturbations of the state lead to relations between the modular Hamiltonian and the emergent geometry that can be interpreted as Einstein's equation, as has been suggested in a holographic AdS/CFT context. In doing so we will follow some of the logic in  \cite{graventanglement2} and  \cite{jacobson15}. In particular, we show that ``nonlocal'' perturbations that entangle distant parts of the emergent geometry, similar to the case in ER=EPR, will give rise to what might be understood as a highly quantum wormhole, where spatial curvature generated by (modular) energy, in a manner similar to Einstein's equation, is localized at the wormhole ``mouths."

Our basic strategy is as follows:
\begin{itemize}
\item Decompose Hilbert space into a large number of factors, $\mathcal{H}= \bigotimes_{p}^N\mathcal{H}_p$. Each factor is finite-dimensional.
\item Consider states $|\psi_0\rangle \in \mathcal{H}$ that are ``redundancy-constrained,'' a generalization of states in which the entropy of a region obeys an area law.
\item Use the mutual information between factors $A$ and $B$, $I(A\co B) = S(A) + S(B) - S(AB)$, to define a metric on the graph connecting the factors $\mathcal{H}_p$.
\item Show how to reconstruct smooth, flat geometries from such a graph metric (when it exists).
\item Consider perturbations $|\psi_0\rangle \rightarrow |\psi_0\rangle + |\delta\psi\rangle$, and show these produce local curvature proportional to the local change in entropy.
\item Relate the change in entropy to that in an effective IR field theory, and show how the entanglement first law $\delta S = \delta\langle K\rangle$ (where $K$ is the modular Hamiltonian) implies a geometry/energy relation reminiscent of Einstein's equation.
\end{itemize}
We do not assume any particular Hamiltonian for the quantum dynamics of our state, nor do we explore the emergence of Lorentz invariance or other features necessary to claim we truly have an effective quantum theory of gravity, leaving that for future work.

We begin the paper by reviewing entropy bounds and properties of entanglement for area-law systems in section \ref{sec:arealaw}.
In section \ref{sec:emergentspace} we introduce the notion of redundancy constraint for entanglement structure and show how an approximate sense of geometry for can emerge in such states. In particular, we give a generic outline of the procedure, followed by an example using an area-law state from a gapped system where it is possible to approximately reconstruct space with Euclidean geometry. Then in section \ref{sec:curvature} we discuss the effects of entanglement perturbations in terms of the approximate emergent geometry, and in \ref{sec:gravity} show that an analog of the Einstein's equations can be derived. Finally, in \ref{sec:conclusion} we conclude with a few remarks. 

As this work was being completed we became aware of a paper with related goals \cite{Raasakka:2016uyk}. There are also potential connections with a number of approaches to quantum gravity, including loop quantum gravity \cite{Rovelli:2011eq}, quantum graphity \cite{Konopka:2006hu}, holographic space-time\cite{Banks:2010tj, Banks:2011av}, and random dynamics \cite{Nielsen:2014rfa}; we do not investigate these directly here.

Throughout this paper, we will use $d$ to denote spacetime dimension and $D=d-1$ for spatial dimensionality.
\section{Area-Law Entanglement}\label{sec:arealaw}

\subsection{Gravity and Entropy Bounds}\label{sec:GEB}

The Bekenstein-Hawking entropy of a black hole in 3+1 dimensions is proportional to the area $\area$ of its event horizon, 
\be
S_\text{BH}=\frac{\area}{4G} = 2\pi\frac{\area}{\ell_p^2},
\label{eqn:bhentropy}
\ee 
where we use $\hbar=c=1$ and the reduced Planck length is $\ell_p = \sqrt{8\pi G}$.
At a quick glance this might seem like a surprising result, as the entropy of a classical thermodynamic system is an extensive quantity that scales with volume rather than area. 
What does this imply about the Hilbert space describing the quantum system that is a black hole, or spatial regions more generally?

Consider a fixed lattice of qubits, with a spacing $\ell_0$ and a linear size $r$. The total number of qubits is $n \sim (r/\ell_0)^D$, where $D$ is the dimensionality of space, and the associated dimension of Hilbert space is $N = 2^n$.
If the system is in a (potentially mixed) state with density matrix $\rho$, the von~Neumann entropy is $S=-\Tr \rho\log \rho$.
The maximum entropy of such a system is then $S_\text{max} = \log_2(N) = n$, proportional to the system volume.
We might guess that gravity provides an ultraviolet cutoff that acts similarly to a lattice with $\ell_0 = \ell_p$.
However, Bekenstein argued that the vast majority of the states included in such a calculation are physically unattainable, and that the entropy of a system with mass $M$ and linear size $R$ is bounded by $S\leq 2\pi RM$ \cite{Bekenstein:1980jp,Bekenstein:1993dz}. 
Since a system with $GM>R/2$ undergoes gravitational collapse to a black hole, this suggests that (\ref{eqn:bhentropy}) represents an upper bound on the entropy of any system in spacetime, a constraint known as the holographic bound.
If we were able to construct a higher-entropy state with less energy than a black hole, we could add energy to it and make it collapse into a black hole; but that would represent a decrease in entropy, apparently violating the Second Law. 
The Bousso bound \cite{Bousso:1999xy} provides a covariant version of the holographic bound.
't Hooft and Susskind built on this argument to suggest the holographic principle: in theories with gravity, the total number of true degrees of freedom inside \emph{any} region is proportional to the area of the boundary of that region \cite{'tHooft:1993gx,Susskind:1994vu}, and such a system can be described by a Hilbert space with dimension of approximately 
\be
\text{dim}\,\mathcal{H}\sim e^{S} \sim e^{(r/\ell_p)^{D-1}}.
\ee

Meanwhile, it is now appreciated that area-law behavior for entanglement entropy occurs in a variety of quantum systems, including many non-gravitational condensed-matter examples \cite{Eisert:2008ur}.
Divide space into a region $A$ and its complement $\overline A$.
A quantum state $|\psi\rangle$ is said to obey an area law if the entropy $S_A$ of the reduced density matrix $\rho_A = \Tr_{\overline A} |\psi\rangle\langle\psi|$ satisfies
\begin{equation}
 S_A = \eta \area + \cdots,
 \label{eqn:arealaw}
\end{equation}
where $\area$ is the area of the surface bounding $A$, and $\eta$ is a constant independent of $\area$.
(Here and elsewhere in this paper, entropy equalities should be interpreted as approximations valid in the limit of large system size.)
This behavior is generally expected in low-energy states of quantum field theories with an ultraviolet cutoff \cite{Bombelli:1986rw,Srednicki:1993im} and those of discrete condensed-matter systems with gapped local Hamiltonians ({\it i.e.}, short-range interactions) \cite{wolf}.
In conformal field theories, Ryu and Takayanagi showed that the entanglement entropy of a region was related to area, not of the region itself, but of an extremal surface in a dual bulk geometry \cite{Ryu:2006bv,Hubeny:2007xt,Faulkner:2013ana}.

The existence of an area law does not by itself imply holographic behavior; holography is a statement about the number of degrees of freedom in a region, which is related to the maximum possible entropy, but not directly to the entropy of some specific state as in (\ref{eqn:arealaw}). (The AdS/CFT correspondence is of course holographic on the dual gravity side, but the CFT by itself is not.)
In either a gapped condensed-matter system or a QFT with an ultraviolet cutoff $\ell_0$, we would still expect degrees of freedom to fill the enclosed volume, and the subsystem in $A$ to have $\text{dim}\,\mathcal{H}_A \sim e^{(r/\ell_0)^D}$.

As the UV cutoff length is taken to zero, we find an infinite-dimensional Hilbert space in any QFT, and the entropy of a region of space will generically diverge.
Nevertheless, QFT reasoning can be used to derive a quantum version of the Bousso bound \cite{Strominger:2003br,Bousso:2014sda,Bousso:2014uxa}, by positing that the relevant entropy is not the full entanglement entropy, but the vacuum-subtracted or ``Casini" entropy \cite{Casini:2008cr}.
Given the reduced density matrix $\rho_A$ in some region $A$, and the reduced density matrix $\sigma_A$ that we would obtain had the system been in its vacuum state, the Casini entropy is given by
\be
  \Delta S = S(\rho_A) - S(\sigma_A) = -\Tr \rho_A\log\rho_A + \Tr \sigma_A\log\sigma_A.
\ee
This can be finite even when Hilbert space is infinite-dimensional and the individual entropies $S(\rho_A)$ and $S(\sigma_A)$ are infinite.
This procedure sidesteps the question of whether the true physical Hilbert space is infinite-dimensional (and the holographic entropy bounds refer to entanglement entropy over and above that of the vacuum) or finite-dimensional (and the Casini regularization is just a convenient mathematical trick).

One might imagine being bold and conjecturing not only that there are a finite number of degrees of freedom in any finite region, as holography implies, but also that the holographic bound is not merely an upper limit, but an actual \emph{equality} \cite{Bianchi:2012ev,Cooperman:2013iqr,Myers:2013lva}. 
That is, for any region of spacetime, its associated entanglement entropy obeys an area law (\ref{eqn:arealaw}).
Evidence for this kind of area law, and its relationship to gravity, comes from different considerations.
Jacobson \cite{Jacobson:2012yt} has argued that if UV physics renders entropy finite, then a thermodynamic argument implies the existence of gravity, and also vice-versa.
Lloyd \cite{Lloyd:2012du} has suggested that if each quantum event is associated with a Planck-scale area removed from two-dimensional surfaces in the volume in which the event takes place, then Einstein's equation must hold.

In this paper, we examine quantum states in a finite-dimensional Hilbert space and look for emergent spatial geometries, under the assumptions that distances are determined by mutual information and that ``redundancy constraint,'' which reduces to the usual area-law relationship of the basic form (\ref{eqn:arealaw}), holds when there exists an emergent geometric interpretation of the state. 
The conjecture that arbitrary regions of space are described in quantum gravity by finite-dimensional Hilbert spaces represents a significant departure from our intuition derived from quantum field theory.

We suggest that the emergence of geometry from the entanglement structure of the state can reconcile $\text{dim}\,\mathcal{H}_A \sim e^{(r/\ell_0)^D}$ (degrees of freedom proportional to enclosed volume) with the holographic principle in a simple way: if we were to ``excite'' states in the interior by entangling them with exterior degrees of freedom, the emergent geometry would be dramatically altered so that the system would no longer resemble a smooth background manifold.
In other words, those degrees of freedom are only ``in the interior'' in a geometric sense when they are entangled with their neighbors but not with distant regions, in a way reminiscent of ER=EPR.

\subsection{Area Laws and Graphs}\label{sec:arealaws}

Simply being given a state $|\psi_0\rangle$ in a Hilbert space $\mathcal{H}$ is almost being given no information at all.
Hilbert space has very little structure, and we can always find a basis $\{|\phi_n\rangle\}$ for $\mathcal H$ such that $\langle \psi_0|\phi_1\rangle=1$ and $\langle \psi_0|\phi_{n>1}\rangle = 0$.
To make progress we need some additional data, such as the Hamiltonian or a decomposition of $\mathcal H$ into a tensor product of factors.
In this paper we don't assume any particular Hamiltonian, but begin by looking at states and decompositions that give us a generalization of area-law behavior for entropy.

To get our bearings, we start by considering systems for which we have an assumed notion of space and locality, and states that obey an area law of the form (\ref{eqn:arealaw}), and ask how such behavior can be recovered in a more general context.
Typically such a state $|\psi_0\rangle$ is a low-lying energy state of a gapped local system. Its entanglement structure above a certain scale seems to capture the space on which the Hamiltonian is defined \cite{tensorandgeometry}. 
The entanglement structure of such states is highly constrained.

A remarkable feature of these states is that the entanglement structure above the said scale can be fully characterized once all the mutual information between certain subsystems are known. 
Divide the system into a set of sufficiently large non-overlapping regions $A_p$. 
We can calculate the entropy $S(A_p)$ of each region, as well as the mutual information $I(A_p\co A_q)$ between any two regions. 
The system therefore naturally defines a  weighted graph $G=(V,E)$, on which vertices $V$ are the regions $A_p$, and the edges $E$  between them are weighted by the mutual information (which is manifestly symmetric). .

The mutual information between regions is a measure of how correlated they are. It provides a useful way of characterizing the ``distance'' between such regions because of its relation to correlation functions between operators.
We expect that in the ground state of a field theory, correlators of field operators will decay as exponentials (for massive fields) or power laws (for massless ones).
The mutual information may reflect this behavior, as it provides an upper bound on the correlation function between two operators. 
For the mutual information between regions $A$ and $B$, we have:
\begin{align}
I(A\co B) &= S(\rho_{AB} || \rho_A\otimes \rho_B)\\
&\geq \frac{1}{2}|\rho_{AB} - \rho_A\otimes \rho_B|^2\\
&\geq\frac{\left\{\Tr[(\rho_{AB} - \rho_A\otimes \rho_B)(\mathcal{O}_A\mathcal{O}_B)]\right\}^2}{2\|\mathcal{O}_A\|\,\|\mathcal{O}_B\|^2}\\
&=\frac{(\langle\mathcal{O}_A\mathcal{O}_B\rangle-\langle\mathcal{O}_A\rangle\langle\mathcal{O}_B\rangle)^2}{2\|\mathcal{O}_A\|^2\,\|\mathcal{O}_B\|^2}.
\label{eqn:mutualinfobound}
\end{align}
We therefore choose to concentrate on mutual information as a way of characterizing emergent distance without picking out any preferred set of operators.

Consider grouping a set of non-overlapping subregions $A_p$ into a larger region $\mathbf{B}$, dividing space into $\bold B$ and its complement $\overbar {\mathbf B}$.
Taking advantage of the short-ranged entanglement in such states, the approximate entanglement entropy of ${\bold B}$ can be calculated using the cut function,
\begin{equation}
 S({\bold B}) = \frac 1 2 \sum_{p\in {\bold B}, q\in \overbar{{\bold B}}} I(A_p\co A_q).
 \label{eqn:cutfcn}
\end{equation}
To find the approximate entanglement of region $\bold B$, one simply cuts all edges connecting $\bold B$ and its complement $\overbar{\bold B}$. The entanglement entropy is the sum over all the weights assigned to the cut edges. This is similar to counting the entanglement entropy by the bond cutting in tensor networks, except in this special case where we are content with approximate entanglement entropy for large regions, a simple graph representation is sufficient. Comparatively,  a tensor network that characterizes this state contains far more entanglement information than the simple connectivity captured by the graphs considered here.

\begin{figure}
 \includegraphics[width=0.65\textwidth]{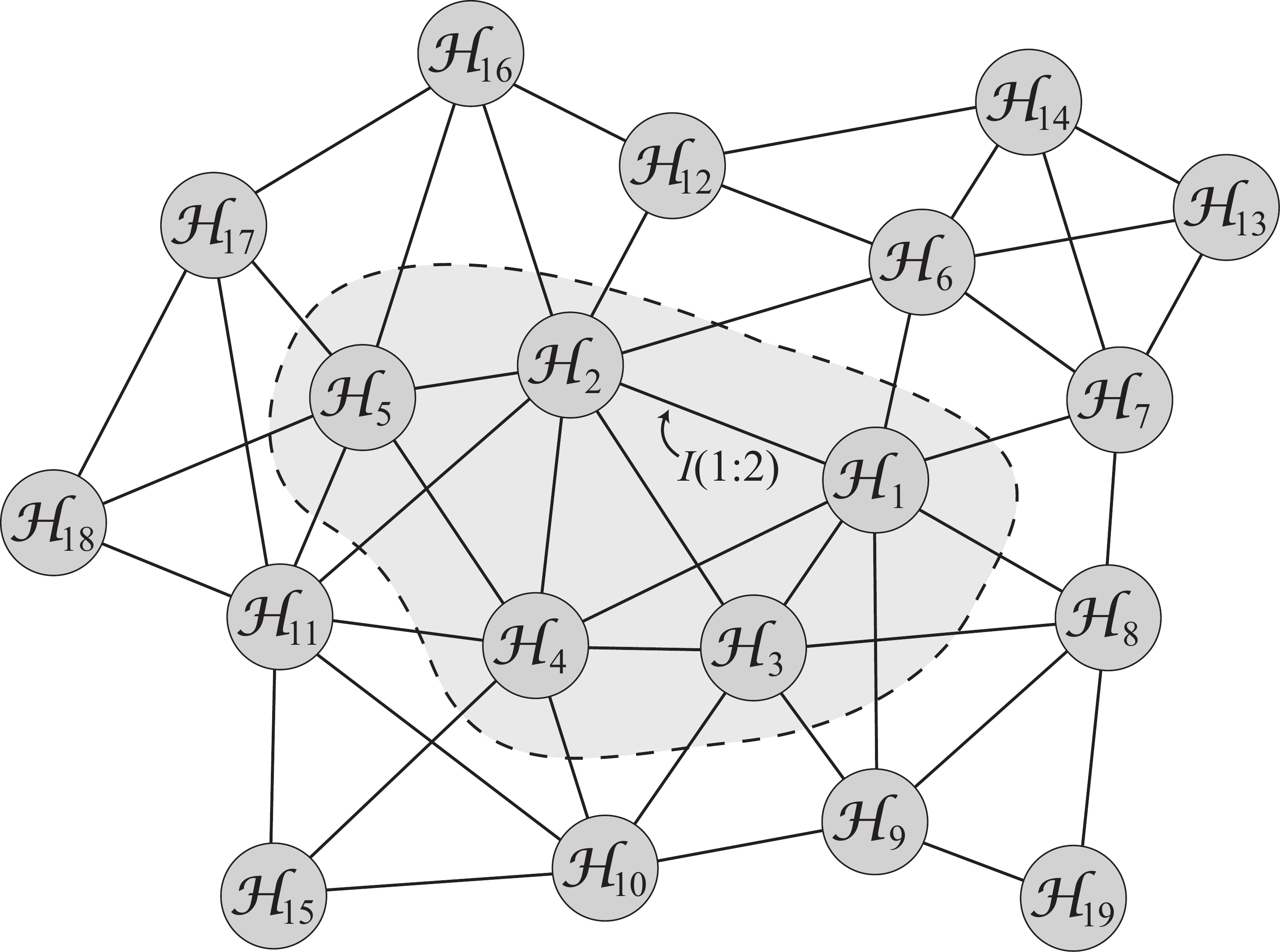}
\caption{An ``information graph'' in which vertices represent factors in a decomposition of Hilbert space, and edges are weighted by the mutual information between the factors. In redundancy-constrained states, the entropy of a group of factors (such as the shaded region ${\bold B}$ containing $\mathcal{H}_1\otimes\mathcal{H}_2\otimes\mathcal{H}_3\otimes\mathcal{H}_4\otimes\mathcal{H}_5$.) can be calculated by summing over the mutual information of the cut edges, as in (\ref{eqn:cutfcn}). In the following section we put a metric on graphs of this form by relating the distance between vertices to the mutual information, in (\ref{eq:weights}) and (\ref{eqn:metric}).
}
\label{fig:regions}
\end{figure}

Our conjecture is that this graph information is enough to capture the coarse geometry of this area-law state. If we restrict ourselves to work at scales for which $S\propto\area$, all information encoded in the form of larger-scale entanglement is highly redundant. In a generic state, the mutual information between all disjoint regions $A_p$, $A_q$ would not be enough to characterize entanglement entropy of $S(A_rA_sA_tA_u)$ for $r,s,t,u\in V$. Naively, to specify the entanglement entropy of all larger regions $\bold B$, one needs on the order of $\mathcal{O}(2^{N})$ data points, where $N$ is the number of vertices (Hilbert-space factors). However, in the special case of area-law entanglement, it suffices to specify all the mutual information between $N$ factors. The amount of classical bits needed to store this is only of order $\mathcal{O}(N^2)$. Therefore, all larger partition entanglement entropy data are ``redundant'' as they are captured by the mutual information of smaller parts. 
Because all subsequent higher-partition entanglement information is encoded in the mutual information between all suitably chosen partitions, the approximate geometric information above the chosen scale of partitioning can be characterized by the graph representation. 

One may worry that the subleading terms in the area-law function can scale as volume and therefore ruin the estimate for higher-partition entropies at some level of coarse-graining. This is, however, an over-estimation.
The entropy of a region with approximate radius $r$ computed by the cut function assumes a strict area law, which scales as $r^{D-1}$ for a $D$-spatial-dimensional area-law system. This is off from the actual entropy by amount $r^{D-2}+\dots$, where missing terms have even lower power in $r$. The relative error, which scales as $r^{-1}$, vanishes in the large-region limit. 

On the other hand, if one keeps all sub-leading terms, the correct edge weights one should assign are given by the intersecting area plus an error term,
\begin{equation}
 I(A_p\co A_q)=\alpha \area(A_p \cap A_q) + \beta \mathcal{E}.
 \label{eqn:info-area}
\end{equation}
 Therefore one may worry that in our subsequent estimate of entropy for a bigger region, the error term may accumulate as $r^{D}$. But since the system is dominated by short-range entanglement, the number of edge cuts only scales as $r^{D-1}$. So in the worst case scenario, the subleading terms will contribute a term that scales as area.
Therefore, the error in using the cut function as an estimate for the entropy of a region $A$ in an area-law system is upper bounded by a term $\beta \area$ for some $\beta$, which one may absorb by redefining $\alpha' = \alpha+\beta$. In this discussion, we are not concerning ourselves with the specific value of $\alpha$, so the sub-leading terms only minimally change the results. 

\section{Emergent Space}\label{sec:emergentspace}

\subsection{Redundancy-Constrained States}\label{sec:RC}

Having established the above properties for area-law states in systems for which space and locality are defined, we now turn to a more general context.
For area-law states, the entanglement information between different factors of Hilbert space is sufficiently redundant that it can be effectively characterized by only limited knowledge of mutual information \cite{arealawcone}. 
In the rest of this work, we restrict ourselves to the study of quantum states that are approximately ``redundancy-constrained,'' defined by slight generalizations of the observations we made for area-law states using purely entanglement information. 

Consider a quantum state $|\psi_0\rangle\in\mathcal{H}=\bigotimes_{p}^N\mathcal{H}_p$. 
We say that the state is \textit{redundancy-constrained} (RC) if, for any subsystem ${\bold B}$ constructed as a tensor product of some subset of the $\{{\mathcal H}_p\}$, its entanglement entropy is given by a cut function of the form (\ref{eqn:cutfcn}), where $A_p$ denotes the subsystem that lives in the Hilbert subspace $\mathcal{H}_p$.
Note that there is no geometric meaning associated with the Hilbert space at this point.

Due to the redundancy of the entanglement entropy information, the entanglement structure for more coarse-grained partitions can be sufficiently captured by quantum mutual information, and hence admit a graph description as in the previous section. The vertices of the graph label subregions, and edge weights are given by their mutual information.
By (\ref{eqn:cutfcn}), it immediately follows that the degree of each vertex $A_p$ (the number of edges emerging from it) is bounded from above by
\be
 \deg(A_p) 
 =\sum_q I(A_p\co A_q) \leq 2S_{max}(A_p)\leq 2\ln D_p,
\ee
where $\dim \mathcal{H}_{A_p} = D_p$.

RC states admit the same graph construction as area-law states, $G=(V,E)$, where vertices are Hilbert-space factors and edges are weighted by the mutual information between them. Such states can be seen as a straightforward generalization of states with area-law scaling that also lie in the area-law entropy cone \cite{arealawcone}. 
As such, they form a superset of area-law states which also satisfy the holographic inequalities  \cite{holocone}. This doesn't imply, however, that such states have holographic duals. It is easy to check that satisfaction of all holographic inequalities is not a sufficient condition to indicate if a state has a holographic dual. (As a simple example, we know that area-law states from a gapped system don't have holographic duals, yet they still satisfy the holographic inequalities.)

The individual Hilbert-space factors $\mathcal{H}_p$ are not necessarily qubits or some other irreducible building blocks of the space.
In particular, they may be further factorizable, and are required to be sufficiently large that redundancy-constraint becomes a good approximation, even if it would not hold at finer scales.
In a phenomenologically relevant model, we would expect each factor to describe not only the geometry but the field content of a region of space somewhat larger than the Planck volume, though we will not discuss those details here.
Note that the RC property is preserved under a coarse-graining operation in which we decompose Hilbert space into factors $\mathcal{H}_P$ that are products of several of the original factors $\mathcal{H}_p$. We discuss coarse-graining more in appendix~\ref{app:coarsegraining}.

RC states are highly non-generic; they represent situations where entanglement is dominated by short-range effects.
For example, a CFT ground state in $D$ dimensions with a holographic dual is not RC, although its entanglement data is still somewhat redundant in that one only needs the entanglement entropy for balls of all radii to reconstruct the AdS geometry \cite{graventanglement1,graventanglement3}. However, additional data encoded in the larger partitions cannot be characterized by mere mutual information between the partitions $A_p$ for any coarse-graining. In this case, the attempt to define entanglement entropy as area or mutual information doesn't quite work in $d$ spatial dimensions any more because there is no simple additive expression for $S(B_X)$ from $I(A_p\co A_q)$. This extra data for larger partitions is essential in constructing the emergent dimension with AdS geometry. 

At the same time, if we have some dual bulk fields living in AdS whose ground state is presumably also short range entangled \cite{qi}, then it may in turn be described by a RC state in AdS with proper coarse-graining. Therefore, if one has the complete holographic dictionary, an experiment of entangling two copies of CFT to create a thermofield double state has its dual experiment with certain constraints in the bulk, where the entanglement now is directly created in the bulk and two copies of AdS are turned into a wormhole. 
Our general program, however, does not rely on the existence of a dual CFT.

Although most states in Hilbert space given a certain decomposition are not redundancy-constrained, RC states seem like an appropriate starting point for investigating theories of quantum gravity, especially if area-law behavior for entropy is universal.
For the remainder of the program, we are going to focus on simple RC states that correspond to flat space in $D$ spatial dimensions.

\subsection{Metric from Information}\label{sec:metric}

Consider a state $|\psi_0\rangle$ for which there exist a decomposition of the Hilbert space such that $|\psi_0\rangle\in\mathcal{H}=\bigotimes_p^N \mathcal{H}_{p}$ is redundancy-constrained. 
Such a state naturally defines a graph $G=(V,E)$, with $N$ vertices labelled by $p$ and each edge $\{p,q\}$ is weighted by the mutual information $I(A_p\co A_q)$.
Without loss of generality, assume $G$ is connected. In the case when $G$ has multiple large disconnected components, one can simply perform the procedure separately for each connected component.

Our conjecture is that this graph contains sufficient information to define another weighted graph, $\tG(\widetilde V, \widetilde E)$, on which the edge weights can be interpreted as distances, thereby defining a metric space. In general, passing from the ``information graph'' $G$ to the ``distance graph'' $\tG$ might be a nontrivial transformation,
\be
  G(V,E) \rightarrow \tG(\widetilde V, \widetilde E),
\ee
with a different set of vertices and edges as well as weights. However, we will make the simplifying assumptions that the vertices and edges remain fixed, so that the graph is merely re-weighted, and furthermore that the distance weight for any edge $w(p,q)$ is determined solely by the corresponding mutual information, $I(A_p\co A_q)$ (where it is nonzero), rather than depending on the rest of the graph.

Our expectation is that nearby parts of space have higher mutual information, while faraway ones have lower.
We therefore take as our ansatz that the distance between entangled factors is some function $\Phi$ of the mutual information, and express this as a new weight $w(p,q)$ on the edges of our graph. That is, for any $p,q\in V$ where $I(A_p\co A_q)\neq 0$, define the edge weights to be
\be
   w(p,q) = \begin{cases}
              \ell_{\mathrm{RC}} \Phi \Big(I(A_p\co A_q)/I_0\Big) ~~~~&(p\neq q)\\
              0 ~~~~&(p=q)
            \end{cases}
	\label{eq:weights}
\ee
for some length scale $\ell_{\mathrm{RC}}$, the ``redundancy-constraint scale." No edges are drawn if $I(A_p\co A_q)=0$. Here we define the ``normalized'' mutual information $i(p:q)\equiv I(A_p\co A_q)/I_0$, where normalization $I_0$ is chosen such that $I(A_p\co A_q)/I_0=1$ when two regions $A_p, A_q$ are maximally entangled. 
In the case when the Hilbert space dimension is constant for all subregions, we have $I_0 = 2S(A_p)_{\text{max}}=2\log (\text{dim}\, \mathcal{H}_D)$. 

The specific form of the scaling function $\Phi$ will presumably be determined by the kind of system we are describing (\emph{e.g.} by the matter content); only some of its basic properties will be crucial to our considerations.
To be consistent with our intuition, we require $\Phi(1)=0$ and $ \lim_{x\rightarrow 0} \Phi(x)=\infty$, namely, the distance is zero when two states are maximally entangled and far apart when they are unentangled. Similar notions were found in \cite{qi,er-eprmvr}. In addition, we choose $\Phi(x)$ to be a non-negative monotonically decreasing function in the interval $[0,1]$, where a smaller mutual information indicates a larger distance. 
For definiteness it may be helpful to imagine that $\Phi(x) = -\log(x)$, as might be expected in the ground state of a gapped system \cite{wolf,hastings}.

We can now construct a metric space in the usual way, treating weights $w(p,q)$ as distances $\tilde{d}(p,q)$.
For vertices connected by more than one edge, the metric $\tilde{d}(p,q)$ is given by the shortest distance connecting $p$ and $q$. Let $P$ be a connected path between $p$ and $q$, denoted by the sequence of vertices $P=(p=p_0, p_1, p_2, \dots p_k=q)$. 
The metric $\tilde{d}(p,q)$ is then
\begin{equation}
 \tilde{d}(p,q) = \min_{P} \{\sum_{n=0}^{k-1} w(p_n,p_{n+1})\}
 \label{eqn:metric}
\end{equation}
for all connected paths $P$.
It is clear from the definition that for a connected component, $\tilde{d}(p,q)=\tilde{d}(q,p)$, $\tilde{d}(p,q)=0 \Leftrightarrow p=q$, and the triangle inequality is satisfied. 

Given a graph with $N$ vertices with a metric defined on it, we would like to ask whether it approximates a smooth manifold of dimension $D \ll N$. Clearly that will be true for some graphs, but not all. One approach is to consider an $r$-ball centered at $p$ using the metric $\tilde{d}$, and compute the entropy of the reduced density matrix obtained by tracing out all regions outside the ball. The fractal dimension near some vertex $p$ can be recovered if
\begin{equation}
S(r, p)\sim r^{D_f}.
\label{eqn:HausdorffDim}
\end{equation}
In general this expression may not converge to an integer $D_f$. In the case of integer dimension, one can then proceed to find a  $D=D_f+1$ dimensional manifold on which $G$ can be embedded that comes closest to preserving the metric $\tilde{d}(p,q)$. We also assign the \textit{interface area} between two subregions $A_p$, $A_q$ as 
\begin{equation}
 \area :=I(A_p\co A_q)/2\alpha,
 \label{eqn:info-alpha2}
\end{equation}
for some constant $\alpha$. Note that this implies the area that encloses the region $A_p$ in a redundancy-constrained state is given by $\area(A_p)=S(A_p)/\alpha$.
We define this to be the emergent spatial geometry of the state and assign geometric labels to the Hilbert space factors based on the embedding. For simple geometries, we will show in section \ref{subsec:mds} that one can use the so-called dimensionality reduction techniques in manifold learning.

\subsection{Classical Multidimensional Scaling}\label{subsec:mds}

We now turn to the problem of going from a graph with a metric to a smooth manifold. One approach is to use Regge calculus, which we investigate in appendix~\ref{app:coarse}. Here we look at an alternative procedure, multidimensional scaling (MDS). For a more detailed review, see e.g. \cite{mdsbook}.

This procedure defines an embedding of the graph into a symmetric manifold; for simplicity, we restrict our attention to cases where the manifold is Euclidean. The embedding is an isometry when the graph itself is exactly flat, but also works to find approximate embeddings for spaces with some small distortion. In our current program, one expects that there exists some natural number $D\ll N$ where the corresponding embedding in $D$-Euclidean space is (approximately) isometric, but there can be distortion since there is some arbitrariness in our choice of the distance function $\Phi$ appearing in (\ref{eq:weights}). 

Consider the distance graph $\widetilde{G} = (V,E)$, with edges weighted by the metric distance $\tilde{d}(p,q)$. These vertices and distances now define a metric space $(V,\tilde{d})$.
The first thing we can do is define the emergent dimension of this discrete space. 
Consider a subset of vertices, $X=\{v_0, v_1,\dots, v_r\} \subseteq G$, equipped with its induced metric.
$X$ is a metric subspace and a $r$-simplex of $V$. Now construct the matrix 
\begin{equation}
 R_{ij} = \frac 1 2 (\tilde{d}(v_i,v_0)^2+\tilde{d}(v_j,v_0)^2-\tilde{d}(v_i,v_j)^2).
\end{equation}
Since the determinant $det(R)=R(v_0,v_1,\dots, v_r)$ is a symmetric function, define simplicial volume 
\begin{equation}
 vol_r(X)=\frac{1}{r!}\sqrt{det(R)},
 \label{eqn:simplicialvolume}
\end{equation}
which is nothing but the spatial volume of the $r$-simplex if $X$ is a subset of Euclidean space equipped with the induced Euclidean metric. The dimension of a metric space, if it exists, is the largest natural number $k$ for which there exists a $D$-simplex with positive volume. 
As demonstrated by  \cite{morgan74}, the metric space can be isometrically embedded into Euclidean space with dimension $d$ if and only if the metric space is flat and has dimension $\leq D$. 

The output of MDS applied to $N$ vertices with distances $\tilde{d}(p,q)$ embedded into a $D$-dimensional space is an $N\times D$ matrix $\mathbf{X}$, which can be thought of as the embedded coordinates of all the vertices: the $n$th row contains the $D$ coordinate values of the $n$th vertex, up to isometric transformations.

To see how this might work, imagine for the moment working backwards: given some coordinate matrix $\mathbf{X}$, how is it related to the distances $\tilde{d}(p,q)$?
First define an $n\times n$ matrix $\mathbf{B} = \mathbf{X X}^t = (\mathbf{XO})(\mathbf{XO})^t$, which is equivalent for coordinate matrix $\mathbf{X}$ up to some arbitrary orthonormal transformation $\mathbf{O}$.
Then we notice that the Euclidean distances between two rows of $\mathbf{X}$ can be written as 
\begin{align}
 \tilde{d}(p,q)^2 &=\sum_{r=1}^d (X_{pr}-X_{qr})^2\\
  &=\sum_{r=1}^d [X_{pr}X_{pr}+X_{qr}X_{qr}-2X_{pr}X_{qr}]\\
 &= B_{pp}+B_{qq}-2B_{pq}.
\end{align}
Therefore, if $\mathbf{B}$ can be recovered only from the Euclidean distances $\tilde{d}(p,q)$, a solution for $\mathbf{X}$ can be obtained. 

The solution $\mathbf{X}$ for the embedding coordinates is non-unique up to isometric transformations. To get a unique solution, we first impose the following constraints such that the embedding is centered at the origin, 
\begin{equation}
 \sum_{p=1}^N X_{pr}=0, ~~~\forall r.
\end{equation}
Then it follows that $\sum_{q=1}^N B_{pq}=0$ and 
\begin{align}
 B_{pq}=-\frac 1 2 \Big(\tilde{d}(p,q)^2 - \frac{1}{N}\sum_{l=1}^N \tilde{d}(p,l)^2-\frac 1 N \sum_{l=1}^N \tilde{d}(l,q)^2+\frac{1}{N^2}\sum_{l,m=1}^N \tilde{d}(l,m)^2 \Big).
\end{align}
This defines the components $B_{pq}$ in terms of the graph distances.

We diagonalize $\mathbf{B}$ via 
\be
\mathbf{B} = \mathbf{V\Lambda V}^t.
\label{eq:diagonal}
\ee
Here $\mathbf{\Lambda}$ is a diagonal matrix with diagonal eigenvalues $\lambda_1\geq\lambda_2\geq \dots \lambda_N$ arranged in descending order. In addition, because $\mathbf{B}$ has rank $D$, we choose the $D$ eigenvectors corresponding to the $D$ non-zero eigenvalues. A solution for $\mathbf{X}$ is then  
\be
\mathbf{\tilde{X}}=(\sqrt{\lambda_1}\mathbf{v_1},\dots,\sqrt{\lambda_d}\mathbf{v_d}),
\ee 
which is an isometric embedding of $N$ points into a $D$-dimensional Euclidean space.

For the case where exact embedding is not possible, \emph{i.e.}, the distance function is Euclidean but with some small deviations, there will be $D$ dominant positive eigenvalues followed by smaller non-zero eigenvalues. We consider the $D$-dimensional embedding to be approximately valid if $\epsilon_{D}=1-\sum_{i=1}^D|\lambda_i|/\sum_{i=1}^N|\lambda_i|$ is sufficiently small.
One can quantify the distortion from exact embedding in various ways. For instance, for the classical MDS algorithm we use here, a so-called stress function is used as a measure of distortion
\begin{equation}
Stress = \sqrt{1-\frac{(\sum_{p,q} \tilde{d}(p,q) d_E(x_p,x_q))^2}{\sum_{p,q}\tilde{d}(p,q)^2\sum_{p,q}d_E(x_p,x_q)^2}},
\end{equation}
where $d_E(x_p,x_q)$ is the Euclidean distance between the two corresponding vertices $p,q$ in the embedded space.
Essentially, MDS analytically generates a set of embedding coordinates in a lower dimensional Euclidean space  \cite{cmds,mdsbook}, where the algorithm seeks an optimal Euclidean embedding such that the inter-vertex distances are best preserved in sense that $Stress$ is minimized. 
Although it also works well for graph embedding in highly symmetric surfaces (hyperbolic and spherical as well as flat)  \cite{cmds,walter,eladkimmel}, it is considered a difficult problem to find an embedding for generic curved manifolds. The matching problem, although non-trivial, can be significantly simplified if the embedding manifold is known \cite{gmds}.

\subsection{Examples with Area-Law States}\label{subsec:examples}

Let's see how MDS works in practice for our redundancy-constrained quantum states. To do this, we will examine states whose geometric interpretation is known, and show that our procedure can recover that geometry.

We start by imagining that an unsuspecting group of theorists have been handed the state $|\psi_0\rangle\in \mathcal{H}=\bigotimes_{p}^N\mathcal{H}_p$, which is actually the ground state of a $d$-dimensional gapped local Hamiltonian that also satisfies an area law. Although the theorists are only given $|\psi_0\rangle$ and its Hilbert space decomposition, they are tasked with finding an approximate geometry for the state. 

We start by constructing the graph $\tG=(V,E)$, where the vertices are labelled by subregions $A_p$, and edge weights are given by the distance function $w(p,q) = \ell_{\mathrm{RC}}\Phi[I(A_p\co A_q)/I_0]$, as in (\ref{eq:weights}). For convenience we choose
 $\Phi(x)=-\ln(x)$. This function is chosen as most finitely correlated states have fast decaying correlation which, in the limit of large distances, is exponentially suppressed \cite{wolf}. In particular, this is satisfied for any system with a spectral gap whose observables commute at large distances  \cite{hastings}. The correlation of any state that is locally entangled (finitely correlated states), {\it e.g.}, ones that can be expressed in terms of MPS or PEPS tensor networks, are expected to take this form. 

\begin{figure*}[t!]
    \centering
    \begin{subfigure}[t]{0.45\textwidth}
        \centering
        \includegraphics[width=\linewidth]{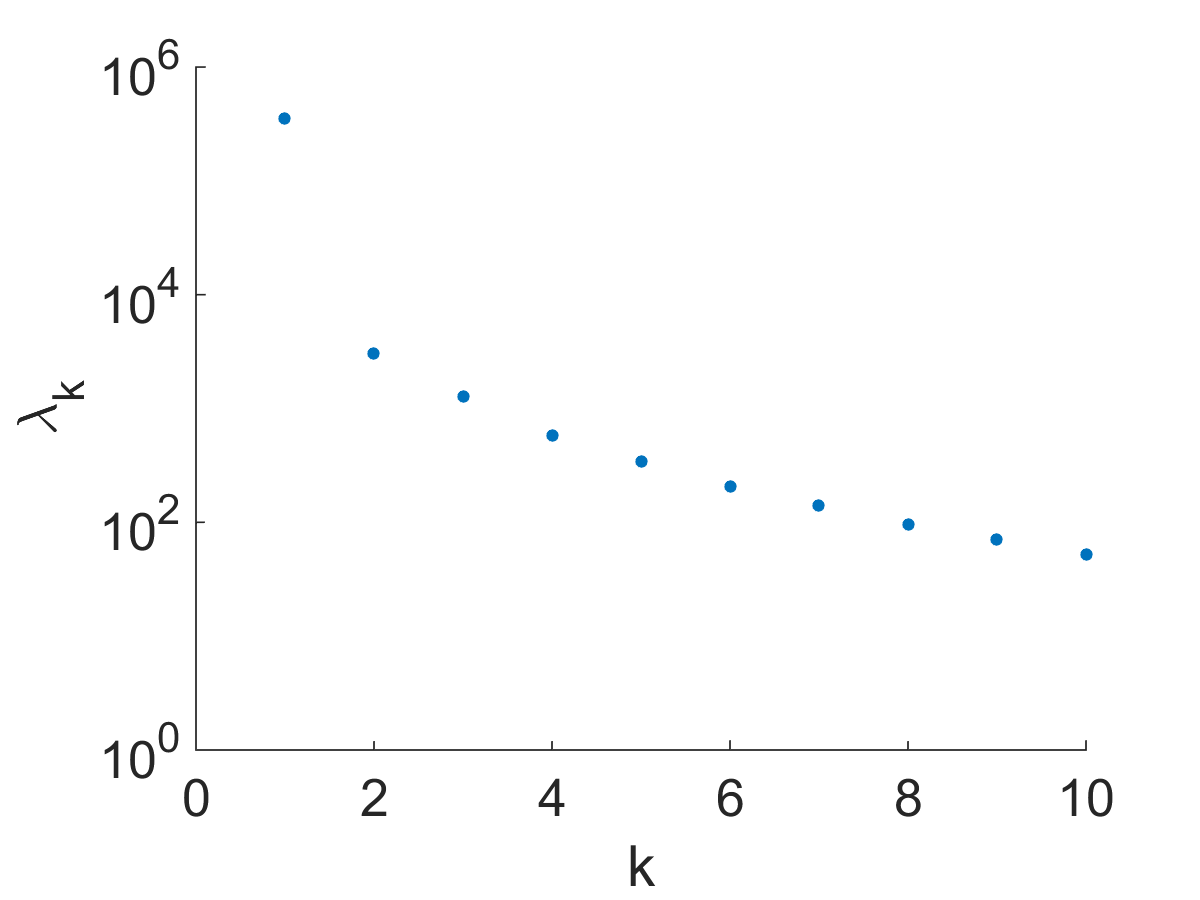}
    \end{subfigure}%
    ~ 
    \begin{subfigure}[t]{0.55\textwidth}
        \centering
        \includegraphics[width=\linewidth]{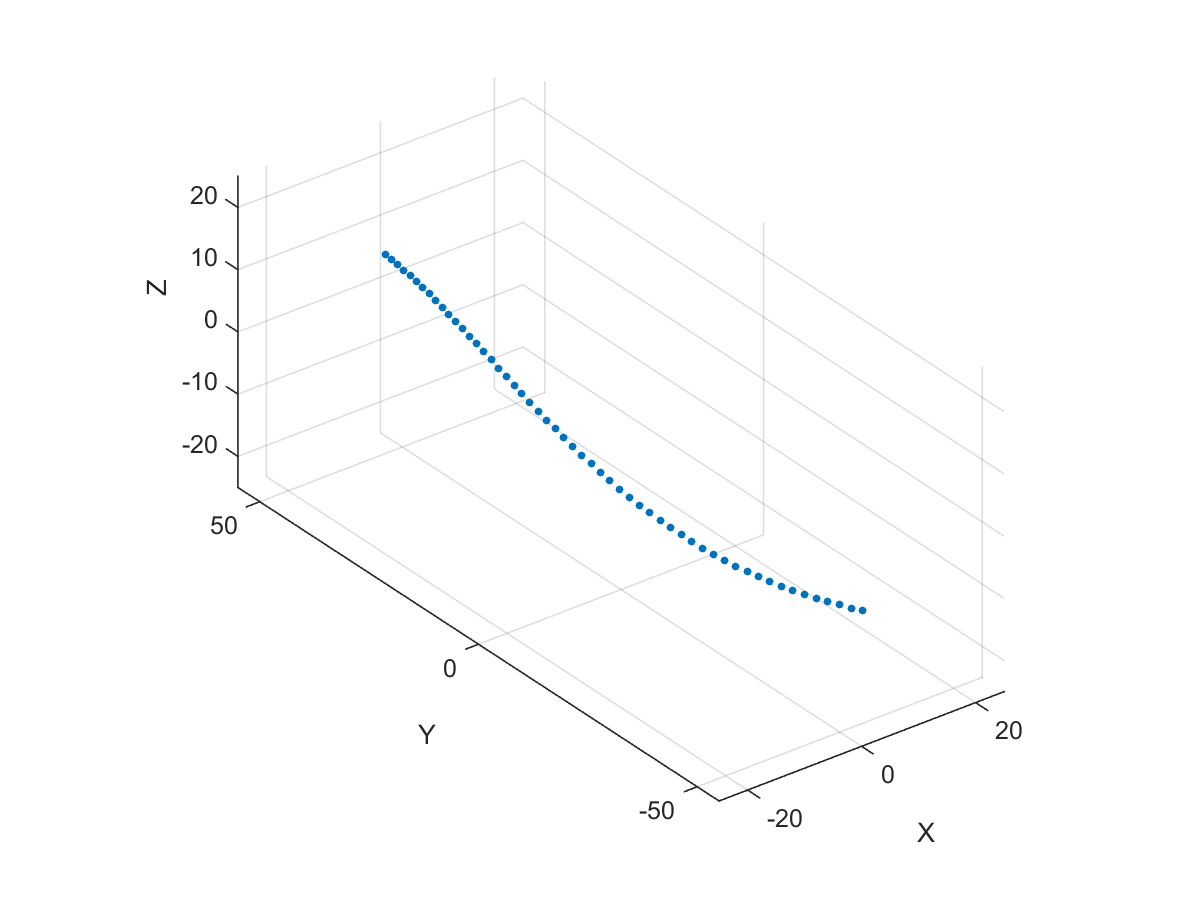}
    \end{subfigure}
  \caption{Multidimensional scaling results for the 1-$d$ antiferromagnetic Heisenberg chain. On the left we plot the eigenvalues of the matrix (\ref{eq:diagonal}). The fact that the first eigenvalue is much greater than the others indicates that we have a 1-$d$ embedding. On the right we show the reconstructed geometry by plotting the first three coordinates of the graph vertices.}
  \label{fig:1dline}
\end{figure*}

Two examples are illustrated here, corresponding to a state living on a one-dimensional line and one living on a two-dimensional plane.
In both cases we start with a known vacuum state of a gapped local system, where correlations are expected be short-ranged.
Computing the mutual information for a quantum state of such systems is in general not an easy task. Consequently, we did not calculate directly the mutual information from density matrices in the 1-$d$ case and instead used the correlation function as a proxy.

Our one-dimensional example is the ground state of an ($S$=1) 1-dimensional antiferromagnetic Heisenberg chain \cite{SorensenAffleck94,Kimetal98}. 
Recalling (\ref{eqn:mutualinfobound}), we use the magnitude squared of the correlation function as an estimate for the mutual information.
The ground state correlation function $|\langle S^a_i S^a_j\rangle|$ is approximately proportional to the modified Bessel function $K_0(r/\xi)$, where $a=x,y,z$. This is a fast-decaying correlation, scaling as as $\exp(-r/\xi)/\sqrt{r/\xi}$ in the asymptotical limit when $r\gg\xi$. (For $a=z$ the correlator is supplemented with an extra term of the same order, given by $\sim 2\xi K_0(r/2\xi)K_1(r/2\xi)/r$. The distortion is still minimal, and in fact yields a slightly better isometric embedding.) 

We constructed a graph of 100 vertices and assigned edge weights given by the square of correlator. No coarse-graining is performed. Applying MDS to this graph returns a vector of embedding coordinates in Euclidean space. As expected, there is distortion in the embedding and the coordinate matrix $\mathbf{\tilde{X}}$ has rank greater than one. However, the distortion, as measured using eigenvalues, has $\epsilon_{1} = 0.0167$, from which we determine that it only slightly deviates from an embedding in 1d. In figure \ref{fig:1dline}, a patch of approximately 50 points is plotted. The one-dimensional nature of the reconstructed geometry is evident.

\begin{figure*}[t!]
    \centering
    \begin{subfigure}[t]{0.45\textwidth}
        \centering
        \includegraphics[width=\linewidth]{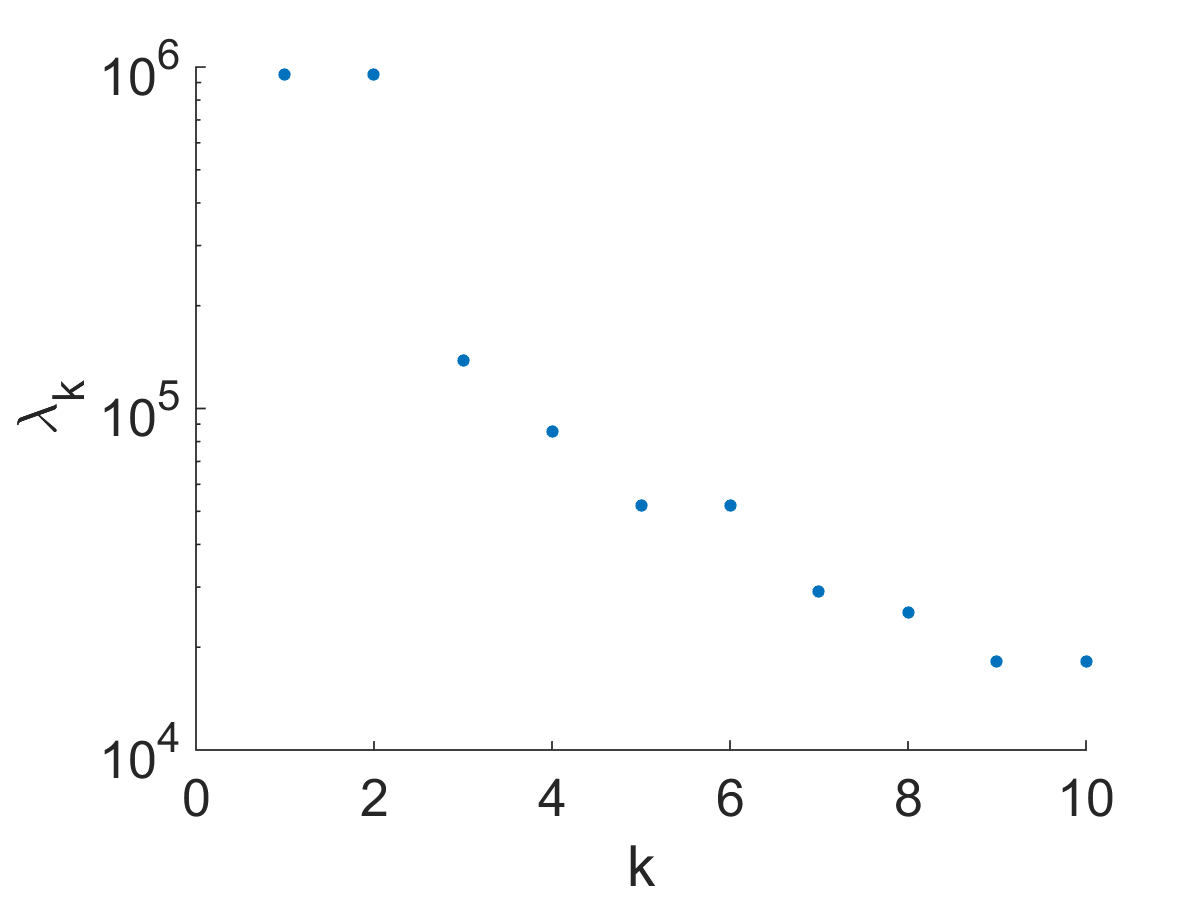}
    \end{subfigure}%
    ~ 
    \begin{subfigure}[t]{0.55\textwidth}
        \centering
        \includegraphics[width=\linewidth]{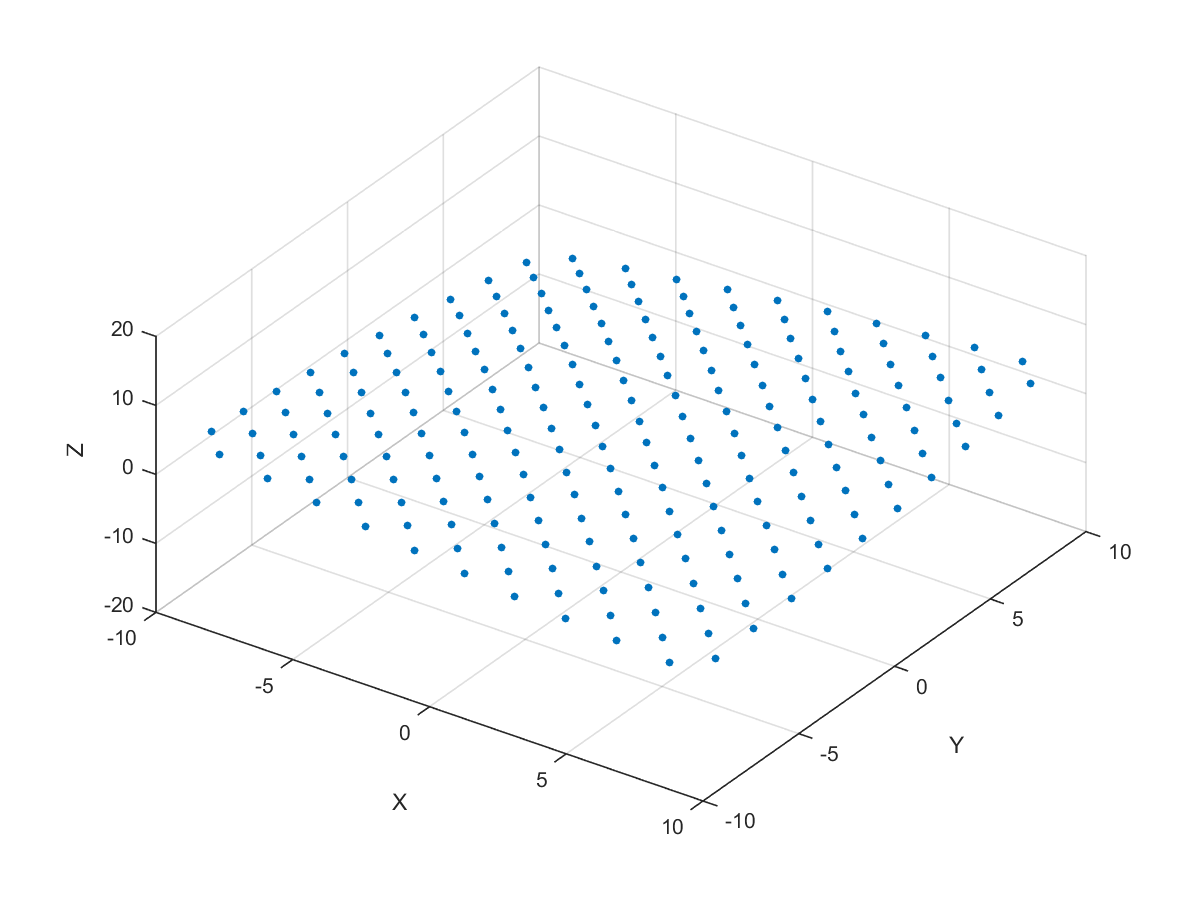}
    \end{subfigure}
    \caption{Multidimensional scaling results for a coarse-grained 2-d toric code ground state. Again, the left shows the eigenvalues of (\ref{eq:diagonal}) and the right shows the reconstructed geometry. The two dominant eigenvalues show that the geometry is two-dimensional, though the fit is not as close as it was in the 1-$d$ example. Similarly, the reconstructed geometry shows a bit more distortion.}
    \label{fig:2darea2d}
\end{figure*}

Our second example reconstructs a patch of the 2-d toric code \cite{Kitaev97}, where it is possible to exactly calculate the entropy for different subregions \cite{Hamma05}. In a coarse-graining where each region is homeomorphic to a plaquette, the exact entropy is $S=\Sigma_{AB}-1=L_{\partial A}-n_2-2n_3-1=n_1+n_2+n_3-1$, where $L_{\partial A}$ is the length of the boundary that separates bipartitions $A$ and $B$. Here, $n_i$ denotes the number of sites/star operators that have $i$ nearest neighbors in $A$. The mutual information used for the network is again given by the length of overlapping boundary, up to a constant correction term. As neighboring spins are uncorrelated, the constant entanglement entropy offset only changes the overall definition of length scale by a constant factor for a near-uniform coarse-graining. As a result, the geometric reconstruction of a 2-d patch is given by (\ref{fig:2darea2d}), up to coordinate rescaling.  
The distortion now is visibly higher because $-\ln(x)$ is no longer an ideal ansatz for $\Phi(x)$. The distortion factor as measured by eigenvalues for an embedding in 2-d has $\epsilon_2 =0.41$, but the 2-d nature of the emergent space is evident from figure~\ref{fig:2darea2d}.

\section{Curvature and Entanglement Perturbations}\label{sec:curvature}

In this section we examine the effects on our reconstructed spatial geometries of perturbing the entanglement structure of our states. 
As we are only considering space rather than spacetime, we cannot directly make contact with general relativity; in particular, we can say nothing about the emergence of dynamical fields obeying local Lorentz invariance.
Nevertheless, we will see that the induced geometry responds to perturbations in a way reminiscent of Einstein's equation, suggesting that an emergent spacetime geometry could naturally recover gravity in the infrared.

\subsection{Entanglement Perturbations}

Consider some unperturbed ``vacuum'' density operator $\sigma=|\psi_0\rangle\langle\psi_0|\in L(\mathcal{H})$, for which there exists a $D$-dimensional geometric reconstruction as discussed in the last section. (Here $L(\mathcal{H})$ denotes the space of complex-valued linear operators on $\mathcal{H}$, of which the density operator is an element.) 
We choose a vertex $p$ on the distance-weighted graph $\widetilde{G}$ that is associated with some subregion $A_p$ of the emergent geometry.
The reduced density matrix associated with the region is defined in the usual way: $\sigma_{A_p} = \Tr_{\overline{A}_p}[\sigma]$, where $\overline{A}_p$ is the complement of $A_p$. 
The entropy of such a region is again $S_{A_p}(\sigma_{A_p})=-\Tr[\sigma_{A_p}\log \sigma_{A_p}]=1/2 \sum I(A_p\co \overline{A}_p)$. The interface area between regions $A_p$ and $A_q$ is defined as $\alpha\area =I(A_p\co A_q)/2$, and the distance between vertices  $p,q$ is defined by $\tilde{d}(p,q)=l_p\Phi(i(A_p\co A_q))$. 
Recall that the normalized mutual information is $i(p\co q)=I(A_p\co A_q)/I_0(p\co q)$, where $i(p\co q)=1$ when subsystems $A_p, A_q$ are maximally entangled. 

There are a variety of entanglement perturbations one can consider. 
We can separately investigate ``local'' perturbations that change the entanglement between $A_p$ and nearby degrees of freedom, and ``nonlocal'' ones that introduce entanglement between $A_p$ and degrees of freedom far away; the latter can be modeled by nonunitary transformations on $\mathcal{H}_{A_p}$. 

A local perturbation is generated by some unitary operator $U_{A_p \overline{A}_p}$ acting on the original system $\mathcal{H} = \mathcal{H}_{A_p}\otimes \mathcal{H}_{\overline{A}_p}$. The perturbed state is $\rho=U^{\dagger}_{A_p \overline{A}_p}\sigma U_{A_p \overline{A}_p}$.
From the definition of mutual information, we know that 
\begin{align}
\delta I({A_p\co  \overline{A}_p}) &= \delta S_{A_p}+\delta S_{\overline{A}_p} -\delta S_{A_p \overline{A}_p}\cr
&=2\delta S_{A_p}, \qquad \mathrm{(local)}
\label{eqn:entropyperturbation}
\end{align} 
where $\delta S_{i}=S_{i}(\rho)-S_{i}(\sigma)$ denotes the infinitesimal change of entanglement entropy for region $i$. (This relation also holds for finite changes in entropy.)
The second equality follows because $\delta S_{A_p}=\delta S_{\overline{A}_p}$ and $\delta S_{A_p \overline{A}_p} =0$, since $U_{A_p \overline{A}_p}$ does not change the total entropy of the system.
By construction, the definitions of area and length are related to mutual information of the quantum state; as we will soon discover, the entanglement perturbation here is tantamount to a local curvature perturbation at $A_p$. 

Nonlocal entanglement perturbations correspond to applying a lossy quantum channel $\Lambda$, which can equivalently be treated as a completely positive and trace preserving (CPTP) map, to the system. 
To that end we introduce an extended Hilbert space
\be
  \mathcal{H}^* = \mathcal{H}_{A_p}\otimes \mathcal{H}_{\overline{A}_p} \otimes \mathcal{H}_B,
  \label{eq:extendedH}
\ee
where $B$ represents some ancillary degrees of freedom that are initially unentangled with those in  $A_p$. 
One can think of $B$, described by state $\sigma_B\in L(\mathcal{H}_B)$, as a different patch of emergent space, or simply some degrees of freedom that the system has not yet encountered. 
A nonlocal perturbation is enacted by a unitary $U_{A_p \overline{A}_p B}= U_{A_p B}\otimes I_{\overline{A}_p}$ that acts only on the degrees of freedom in $A_p$ and $B$.
The perturbed state is $\rho = \Lambda (\sigma)=\Tr_{B}[U_{A_p \overline{A}_p B}^{\dagger}\sigma\otimes\sigma_BU_{A_p \overline{A}_p B}]$. In this case, the change in mutual information between $A_p$ and $\overline{A}_p$ is non-positive, and will depend on the local entanglement structure as well as the entangling unitary. 

Let $F_{\Lambda}(\Delta S_{A_p}; A_p, \overline{A}_p)$ be a function that describes the finite change in mutual information between $A_p$ and its complement. The specific implementation of this function will depend on $\Lambda$ and the entanglement structures related to the regions of interest, $A_p$ and $\overline{A}_p$. The total change in mutual information in this case is given by
\begin{equation}
\Delta I(A_p:\overline{A}_p) = F_{\Lambda}(\Delta S_{A_p};A_p,\overline{A}_{p}).
\end{equation}
Because the change in mutual information has to be zero when no unitary is applied, we must have $F_{\Lambda}(0; A_p, \overline{A}_p)=0$. 
For infinitesimal perturbations, we can write
\begin{align}
\delta I(A_p\co \overline{A}_p) &=\delta S_{A_p}\frac{dF_{\Lambda}(\Delta S_{A_p}; A_{p},\overline{A}_p)}{d(\Delta S_{A_p})}\Big|_{\Delta S_{A_p}=0}\\
&=\delta S_{A_p}F_{\Lambda}'(0;A_p,\overline{A}_p). \qquad \mathrm{(nonlocal)}
\label{eqn:nonlocal}
\end{align}
We have written the change in mutual information as if it is proportional to the change in entropy, but note that (in contrast with the case of local perturbations) here the proportionality is not a universal constant, but rather a factor that depends on the channel $\Lambda$.
 In general, for mixed states $\sigma, \sigma_B$ which are not maximally entangled, we can easily find unitary operations where $F_{\Lambda}'(0;A_p,\overline{A}_p)\neq 0$. 
Note that this relation differs from the local perturbation by an $F'_{\Lambda}$-dependent constant factor.

\subsection{Geometric Implications}

We now consider the effect of an entanglement perturbation on the emergent spatial geometry.
In this section we imagine mapping our graph to a Riemannian embedding manifold, as we did for the vacuum case using MDS in the previous section.
In appendix~\ref{app:coarse} we study the problem using Regge calculus.

Although it is operationally difficult to find for the graph an embedding manifold with variable curvature, it is considerably more tractable if we only wish to quantify a perturbation around a known manifold that corresponds to the density matrix $\sigma$. Namely, in order for the manifold to be a good embedding, its perturbed form should at least be consistent with the deviations in area and geodesic lengths. Since we have outlined explicit algorithms for flat space configurations, here we assume that a $D$-dimensional flat configuration $\mathcal{M}$ has been obtained using the above framework.

\subsubsection{Effects of Local Entanglement Perturbations}

We begin by considering a local perturbation that decreases the entropy of our region, 
\be
\delta S_{A_p}<0, 
\ee
which as we will see induces positive spatial curvature. Thus we are considering a local operation that decreases the entanglement between $A_p$ and the rest of Hilbert space.
Without altering the dimension, the minimal change to $\mathcal{M}$ that can be imposed is some perturbation to spatial curvature at $p$. For the simplest case, let $A_p$ be a region that contains a single graph vertex $p$ whose entanglement with the adjacent vertices $q$ (regions $A_q$) is gradually decreased.

\begin{figure}
 \includegraphics[width=0.65\textwidth]{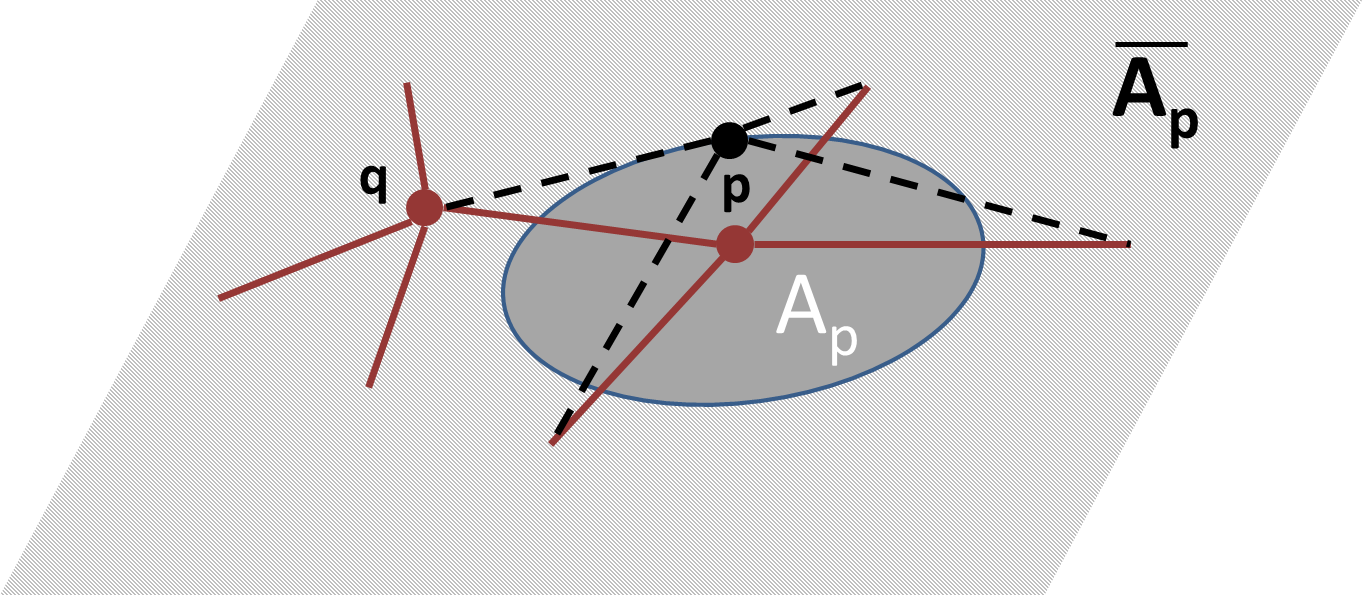}
\caption{For a graph $\widetilde{G}$ embedded in some manifold, we assign subregion $A_p$ (dark blue region) to the vertex $p$, which is connected to adjacent vertices $q$ (Black solid line). An entanglement perturbation that decreases the mutual information between $A_p$ with its neighbors elongates the connected edges (dashed red lines), creating an angular deficit which is related to the curvature perturbation at $p$. 
}
\label{fig:regions}
\end{figure}

Following Jacobson \cite{jacobson15}, we proceed by defining Riemann normal coordinates in the vicinity of $A_p$, 
\begin{equation}
 h_{ij} =\delta_{ij} -\frac{1}{3}r^2 R_{ijkl} x^k x^l + O(r^3).
\end{equation}
Consider the perturbed subregion fixed at some constant volume $V$, of characteristic linear size $r=V^{1/D}$. The decrease in area under the perturbation is given by 
\begin{equation}
 \delta \area = -\frac{\Omega_{D-1}r^{D+1}}{2D(D+2)} \mathcal{R}_p,
 \label{eqn:delta-A-R}
\end{equation}
where $\mathcal{R}_p = R_{ij}{}^{ij}(p)$ is the spatial curvature scalar and $\Omega_{D-1}$ is the volume of a unit $(D-1)$-sphere.
We know that the boundary area is defined by local mutual information, namely,
\begin{equation}
\delta \area = \frac{1}{2\alpha}\sum_{A_q\in \overline{A}_p}\delta I(A_p\co A_q)\approx\frac{1}{2\alpha} \delta I(A_p\co \overline{A}_p).
\label{eqn:areamutualinfo}
\end{equation}
Because the system is only short-range entangled, the local mutual information is well-approximated by the mutual information between the region and its complement. For instance, if the graph that captures entanglement structure for the toric code ground state above the RC scale is used, then the two quantities will be exactly equal. In general for systems with exponentially decaying mutual information, the error with this estimation is also upper-bounded by a quantity of order $\exp(-r/\ell_{\mathrm{RC}})$, which vanishes as long as the vertices correspond to sufficiently coarse-grained regions. 

Plugging (\ref{eqn:entropyperturbation}) for the change in mutual information due to an infinitesimal local perturbation into (\ref{eqn:delta-A-R}) and (\ref{eqn:areamutualinfo}), we can relate the curvature scalar to the entropy perturbation by 
\begin{equation}
\mathcal{R}_p=-\frac{2D(D+2)} {\alpha\Omega_{D-1}(\gamma \ell_{\mathrm{RC}})^{D+1}}\delta S_{A_p}.
\label{eqn:ctscurvature}
\end{equation}
Here we have set $r$ equal to $\gamma \ell_{\mathrm{RC}}$, a characteristic size of the region for some constant $\gamma$. 
Such an approximation is most accurate when the symmetry is also approximately reflected by the graph.
Because $\delta S_{A_p} < 0$, the induced curvature is positive.

The relation between curvature and the entropy perturbation can alternatively be derived by using (\ref{eqn:areamutualinfo}) to estimate the decrease in mutual information for each individual edge, and relate that to a length excess. Since the edge weights sum to the total change in entropy, for each edge we can write 
\be
\delta I(A_p\co A_q)=2\eta_q \delta S_{A_p},
\ee
for some constants $\eta_q$ such that $\sum_q \eta_q=1$. The values of $\eta_q$ are determined by the graph structure near $p$, as well as the unitary $U_{A_p\overline{A}_p}$. For a unitary that symmetrically disentangles all the edges on a regular lattice, $\eta_q=1/\deg(p)$. 

Alternatively, we can also relate the change in entropy to the curvature perturbation by considering the change in linear size.
The radius excess for the same perturbation at some fixed area is
\begin{equation}
 \delta d = \frac{r^3 \mathcal{R}_p}{6D(D+2)}.
 \label{eqn:delta-d}
\end{equation}
Recall that distance is related to mutual information by $\tilde{d}(p,q) = \ell_{\mathrm{RC}}\Phi(i(p\co q))$, where the normalized mutual information is $i(p\co q) = I(A_p \co A_q)/I_0(p:q)$. 
Assuming an approximately symmetric configuration, to leading order we have
\begin{align}
\label{eqn:distancemutualinfo}
\delta \tilde{d}(p,q)&= \ell_{\mathrm{RC}}\Phi' (i(p\co q))\delta i(p\co q) +O(\delta i^2)\\ \nonumber
&=-2\eta_q \ell_{\mathrm{RC}}|\Phi'(i(p\co q))|\delta S_A/I_0(p\co q)+O(\delta i^2),
\end{align}
where for the last line we used (\ref{eqn:entropyperturbation}) to relate the linear change in entropy to the change in the distance function. Note that $\Phi' = d\Phi/di$ is negative by construction.

Comparing (\ref{eqn:delta-d}) to (\ref{eqn:distancemutualinfo}) yields an alternative relation between curvature and entanglement entropy, 
\begin{equation}
 \mathcal{R}_p = -\frac{12\eta_q |\Phi'(i(p\co q))| D(D+2)}{I_0(p\co q)\ell_{\mathrm{RC}}^2\gamma^3} \delta S_{A_p},
\label{eqn:curvatureLengthExcess}
\end{equation}
where again we have set $r=\gamma \ell_{\mathrm{RC}}$.
One can check that for $1/\alpha \propto \ell_{\mathrm{RC}}^{D-1}$, the two results (\ref{eqn:ctscurvature}) and (\ref{eqn:curvatureLengthExcess}) are equivalent up to some dimension-dependent choice of $\eta_q$, the inverse function $\Phi$, and constant factor $\alpha$.
Both imply positive curvature for local disentangling perturbations. Similarly, entangling perturbations with $\delta S_{A_p}>0$ yields negative spatial curvature.

\subsubsection{Effects of Nonlocal Entanglement Perturbations}

The derivation with nonlocal entanglement perturbation is similar, where we simply replace the constant proportionality factors with the channel-dependent factor $F'_{\Lambda}(0)$. Repeating the above steps, analogously to the area deficit condition (\ref{eqn:ctscurvature}) we have 
\begin{equation}
\mathcal{R}_p = -\frac{D(D+2)F_{\Lambda}'(0;A_p,\overline{A}_p)}{\alpha\Omega_{D-1}(\gamma \ell_{\mathrm{RC}})^{D+1}}\delta S_{A_p},
\end{equation}
while analogously to the radius deficit (\ref{eqn:curvatureLengthExcess}) we obtain
\begin{equation}
\mathcal{R}_p=-\frac{6F_{\Lambda}'(0;A_p,A_q)|\Phi'(i(p:q))|D(D+2)}{I_0(p:q)\ell_{\mathrm{RC}}^2\gamma^3}\delta S_{A_p}.
\end{equation}

Interestingly, we find that nonlocal entanglement perturbations are only able to generate positive curvature perturbations.
Because $\delta I(A_p:\overline{A}_p)\leq 0$ under any operations acting on $A_p$ and $B$, from (\ref{eqn:nonlocal}) we must have 
\be
F'_{\Lambda}(0;A_p,\overline{A}_p)< 0
\label{eqn:fprime}
\ee
for a generic entangling unitary when $\delta S_A>0$. If $\delta S_{A_p}<0$, then it must follow that $F'_{\Lambda}(0; A_p,\overline{A}_p)>0$ for the same reason. 

The nonlocal case is also interesting due to its connection with the ER=EPR conjecture. In this case, some spatial region $A_p$, described by some mixed state $\sigma_{A_p}$, is far separated from some other spatial region $B$, with corresponding mixed state $\sigma_B$. An entangling unitary then creates some weak entanglement between the regions, similar to having an EPR pair shared between them. Such entanglement, as we have seen, decreases the mutual information between $A_p$ ($B$) and their respective neighboring regions. This, we claim, can be interpreted as a quantum proto-wormhole. No smooth classical geometry is present to form the usual ER bridge; nevertheless, the entanglement backreacts on the emergent geometry in a way such that positive modular energy ``curves'' the spatial regions near the ``wormhole mouths." 

\begin{figure}
\includegraphics[width=0.7\linewidth]{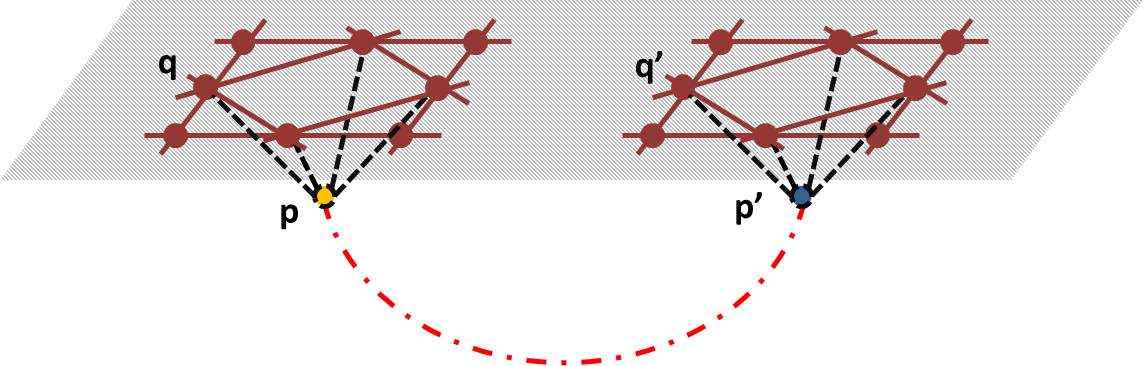}
\caption{For perturbations that slightly entangle two regions of the emergent space, as represented by the vertices, a positive curvature perturbation is induced locally near each perturbed site. We may interpret this as a highly ``quantum wormhole.'' The dotted red line joining $p$ and $p'$ denotes some trace amount of entanglement between the two subsystems. }
\end{figure}

A large entanglement modification beyond the perturbative limit does not create disconnected regions. Heuristically, because the entanglement of $\overline{A}_p$ is always constant under such a unitary, it must become more entangled with region $B$. From the point of view of emergent geometry from entanglement, it implies that the region $\overline{A}_p$ should also be connected to the distant region $B$ in some way. When such a connection between the two regions becomes manifestly geometric, the process may then be interpreted as the formation of a classical wormhole. 

However, because the function $F_{\Lambda}$  depends on the entanglement structure of $\rho_{A\overline{A}}$ and $\Lambda$  in addition to $\Delta S_{A}$, the entropy-curvature relation does not seem to be universal as in the local case. Interpreting nonlocal effects as ``gravitational''  in this model may be in tension with our expectations for a theory of gravity, although further assumptions on symmetries in entanglement structure may resolve this issue. 

\section{Energy and Einstein's equation}
\label{sec:gravity}

We have seen how spatial geometry can emerge from the entanglement structure of a quantum state, and how that geometry changes under perturbations.
This is a long way from completely recovering the curved spacetime of general relativity, both because we don't have a covariant theory with dynamics, and because we haven't related features of the state to an effective stress-energy.
We can address the second of these points by considering general features of a map from our original theory to that of an effective field theory on a fixed spacetime background, then appealing to the entanglement first law (EFL); we leave the issue of dynamics to future work.
Our approach here is similar in spirit to previous work in AdS/CFT  \cite{relativeentropyandholography,graventanglement1,graventanglement2,graventanglement3}
and bulk entropic gravity \cite{jacobson15,Casini:2016rwj,Carroll:2016lku}.

\subsection{Renormalization and the Low Energy Effective Theory}

To make contact with semiclassical gravitation, we need to understand how an effective field theory, and in particular a local energy density, can emerge from our Hilbert-space formalism. This is a nontrivial problem, and here we simply sketch some steps toward a solution, by integrating out ultraviolet (gravitational) degrees of freedom to obtain an infrared field theory propagating on a background.
We argue that, for local entanglement perturbations, a perturbation that decreases the entropy of a quantum-gravity state in a region $A$ will correspond to increased entropy density in the effective IR field theory, which lives in a lower-dimensional (coarse-grained) Hilbert space.

Our construction posits an RC scale $\ell_{\mathrm{RC}}$ at distances greater than which the state obeys the redundancy-constraint condition.
Intuitively we expect $\ell_{\mathrm{RC}}$ to be close to the reduced Planck length $\ell_{p}$, at which the spatial geometry has barely emerged. 
Let the corresponding UV energy scale be $\lambda_{\mathrm{RC}} = 1/\ell_{\mathrm{RC}}$.
We imagine an ``emergence'' map 
\be
\mathcal{E}:\mathcal{H}\rightarrow \mathcal{H}^{\mathrm{EFT}(\lambda_{\mathrm{RC}})},
\ee
which maps states in the Hilbert space of our original theory to those of an effective field theory with cutoff $\lambda_{\mathrm{RC}}$ in a semiclassical spacetime background.

To study the relationship of entropy and energy in this emergent low-energy effective description, we consider the RG flow of the theory, defined by a parameterized map that takes the theory at $\lambda_{\mathrm{RC}}$ and flows it to a lower scale $\lambda$ by integrating out UV degrees of freedom:
\be
\mathcal{F}_\lambda:\mathcal{H}^{\mathrm{EFT}(\lambda_{\mathrm{RC}})}\rightarrow \mathcal{H}^{\mathrm{EFT}(\lambda)}.
\ee
This flow is defined purely in the context of QFT in curved spacetime, so that the background geometry remains fixed. Note that we could also discuss RG flow directly in the entanglement language, where it would be enacted by a tensor network similar to MERA \cite{mera}. There, the equivalent of $\mathcal{F}_\lambda$ would be a quantum channel that could be defined by a unitary circuit if we include ancillae representing UV degrees of freedom that are integrated out. In this case, $\lambda$ depends on the number of layers of the MERA tensor network. 

Given that (\ref{eqn:ctscurvature}) relates a spatial curvature perturbation in the emergent geometry to a change in entropy in the full theory, we would like to know how this entropy change relates to that of the vacuum-subtracted entropy $S^{\mathrm{EFT}(\lambda)}$ in the effective field theory with cutoff $\lambda$ defined on a background (and ultimately to the emergent mass/energy in that theory).
We posit that they are related by a positive constant $\kappa_\lambda$ that depends on the cutoff but is otherwise universal:
\be
  \delta S = -\kappa_\lambda \delta S^{\mathrm{EFT}(\lambda)}.
  \label{eqn:deltaSeft}
\ee

The minus sign deserves some comment. A perturbation that disentangles a Hilbert space factor $A_p$ decreases its entropy, $\delta S_{A_p}< 0$, while inducing positive spatial curvature in the emergent geometry and a decrease in the area $\area_p$ of the boundary of the corresponding region. Naively, a decrease in boundary area results in the decrease of the entropy of a region in a cutoff effective field theory. However, that expectation comes from changing the area of a near-vacuum state in a fixed background geometry. Here we have a different situation, where the perturbation affects both the geometry and the EFT state defined on it. 

In that context, as Jacobson has argued \cite{jacobson15}, we expect an equilibrium condition for entanglement in small regions of spacetime. Consider a perturbation of the EFT defined on a semiclassical background, which changes both the background geometry and the quantum state of the fields. Fix a region in which we keep the spatial volume constant. In the spirit of holography and the Generalized Second Law, the total entropy in the region can be considered as the sum of an area term representing UV quantum gravity modes plus a term for the IR effective field theory,
\be
  \delta_V S_{\mathrm{total}} = \eta\delta_V\area + \delta_VS^{\mathrm{EFT}(\lambda)}.
\ee
Here, the subscript $V$ reminds us that we are considering a variation at fixed volume (in contrast with the original perturbation in our underlying quantum theory). The unperturbed state is taken to be an equilibrium vacuum state.
The total entropy is therefore at an extremal point, $\delta_V S_{\mathrm{total}}=0$.
A decrease in the geometric entropy (represented by the boundary area) is thus compensated by an increase in the entropy of the EFT state. Since our original perturbation $\delta S$ decreases the boundary area of our region, we expect the field-theory entropy to increase. This accounts for the minus sign in (\ref{eqn:deltaSeft}).

Plugging (\ref{eqn:deltaSeft}) into (\ref{eqn:ctscurvature}) produces a relation between the local scalar curvature of a region around $p$ and the change in the entropy of the EFT state on the background:
\begin{equation}
\mathcal{R}_p=\frac{2D(D+2)\kappa_\lambda} {\alpha\Omega_{D-1}(\gamma \ell_{\mathrm{RC}})^{D+1}}\delta S^{\mathrm{EFT}(\lambda)}_{A_p}.
\label{eqn:eft-curvature}
\end{equation}
Here, $\delta S^{\mathrm{EFT}(\lambda)}_{A_p}$ is interpreted as the change in the entropy of the EFT state in the region defined by $A_p$, due to shifts in both the background geometry and the fields themselves. Considering nonlocal rather than local perturbations results in multiplying the right-hand side by $F'_{\Lambda}(0)/2$.

\subsection{Energy and Gravity}

We can now use the Entanglement First Law to relate the change in entropy to an energy density.
The EFL, which relates changes in entropy under small perturbations to the system's modular Hamiltonian,
holds true for general quantum systems. 
Given a density matrix $\sigma$, we define its associated modular Hamiltonian $K(\sigma)$ through the relation
\begin{equation}
\sigma=\frac{e^{-K}}{\Tr(e^{-K})}.
\end{equation}
The Kullback-Leibler divergence, or relative entropy, between two density matrices $\rho$ and $\sigma$ is given by
\begin{align}
D(\rho || \sigma) &\equiv \Tr(\rho\ln\rho)-\Tr(\rho\ln\sigma)\cr
&=- \Delta S + \Delta\langle K(\sigma)\rangle,
\end{align}
where
\be
  \Delta S = S(\rho) - S(\sigma), \quad \Delta \langle K(\sigma)\rangle = \Tr[\rho K(\sigma)] - \Tr[\sigma K(\sigma)].
\ee
The relative entropy is nonnegative, and is only zero when the states are identical. Hence, for infinitesimal perturbations $\sigma = \rho + \delta \rho$, we have $D(\sigma_{}||\rho_{})=0$ to linear order, and we arrive at the EFL \cite{relativeentropyandholography},
\begin{equation}
  \delta S_{} = \delta\langle K_{}\rangle .
  \label{eqn:efl}
\end{equation}
This equation allows us to establish a relationship between (modular) energy and the change in entropy, and thereby geometry, of our emergent space.

Comparing to the curvature-entropy relations for local perturbations (\ref{eqn:ctscurvature}), we see that the (positive) induced local curvature $\mathcal{R}_p$ is proportional to $-\delta \langle K_{A_p}\rangle$. We therefore define an effective modular
energy density,
\begin{equation}
 \varepsilon_p=-\delta \langle K_{A_p}\rangle.
\end{equation}
The curvature is then related to the effective modular energy via
\begin{equation}
 \mathcal{R}_p=-\zeta\delta S_{A_p}=\zeta \varepsilon_p,
\end{equation}
where
\begin{equation}
 \zeta\equiv\frac{2D(D+2)}{\alpha\Omega_{D-1}(\gamma\ell_{RC})^{D+1}}
\end{equation}
is a positive constant. On the other hand, an entangling operation in the vicinity of $A_p$ would give rise to a negative ``energy'' in the region, adding negative spatial curvature to the emergent geometry. 
Having a large amount of negative energy seems unphysical; if our model for emergent space is to be consistent with gravity as we know it, there must be conditions limiting such effects, such as instabilities or other dynamical processes rendering them unattainable. 

For nonlocal perturbations, we saw from (\ref{eqn:fprime}) that only positive curvature is generated, regardless of the precise form of the perturbation. The curvature-modular-energy relation for nonlocal perturbations is therefore
\begin{equation}
 \mathcal{R}_p=\tilde{\zeta}|\varepsilon_p|,
\end{equation}
where $\tilde{\zeta}=|F'_{\Lambda}(0)|\zeta/2$.

Thus, once we define an emergent geometry using mutual information, we see that perturbing the modular energy induces scalar curvature in the surrounding space.
This is manifestly reminiscent of the presence of mass-energy in the region for the usual case in Einstein gravity. 
This relationship between energy and curvature did not come about by ``quantizing gravity''; rather, it is a natural consequence of defining the emergent geometry in terms of entanglement.

However, the effective modular energy is only an analogous expression for the actual mass/energy. To connect with our familiar notion of energy, we need to find the explicit expression of the modular Hamiltonian in terms of a stress tensor. Such expression will generally be highly nonlocal, and is not known explicitly except for a few cases \cite{fredenhagen1985}. 
One exception, however, is for a conformal field theory, where $K$ can be directly related to the stress-energy tensor $T_{\mu\nu}^{\mathrm{CFT}}$ \cite{graventanglement1,relativeentropyandholography}. Consider a small region centered at $p$ of size $\gamma \ell_{\mathrm{RC}}$, in which $T^{\mathrm{CFT}}_{00}$ is approximately constant. Then we have
\begin{equation}
\delta\langle K^{\mathrm{CFT}}_{A_p} \rangle = \frac{2\pi\Omega_{D-1}(\gamma \ell_{\mathrm{RC}})^{D+1}}{D(D+2)}\delta\langle T_{00}^{\mathrm{CFT}}(p)\rangle.
\label{eqn:cftmodularenergy}
\end{equation}
Here $\delta \langle T_{00}^{\mathrm{CFT}}\rangle=\Tr[\delta \rho^{(IR)}_{\lambda} T_{00}^{\mathrm{CFT}}]$ is the expectation with respect to the perturbed state of the IR effective theory. Of course the $00$ component of the stress tensor is simply the energy density of the theory.

Suppose that the RG flow $\mathcal{F}_\lambda$ for our EFT passes through an IR fixed point at a scale $\lambda_*$. At that scale, the entanglement structure of the IR state can be approximated by the ground state of a conformal field theory, and (\ref{eqn:cftmodularenergy}) applies. Combining (\ref{eqn:efl}) and (\ref{eqn:cftmodularenergy}) with (\ref{eqn:eft-curvature}) in the context of of the low-energy effective field theory, we find that the spatial curvature is related to the EFT stress tensor by
\begin{equation}
\mathcal{R} = \frac{4\pi\kappa_{\lambda_*}}{\alpha} \delta\langle T_{00}^{\mathrm{CFT}}\rangle,
\label{eqn:curvaturemodularenergy}
\end{equation}
where $\alpha$ is the constant in the entanglement/area relation (\ref{eqn:info-alpha2}) and $\kappa_{\lambda_*}$ relates the full entropy perturbation to that of the EFT as in (\ref{eqn:deltaSeft}).
For nonlocal perturbations, we saw from (\ref{eqn:fprime}) that only positive curvature is generated, regardless of the precise form of the perturbation.
The curvature-modular-energy relation for nonlocal perturbations is therefore
\begin{equation}
\mathcal{R} = \frac{2\pi|F_{\Lambda}'(0)|\kappa_{\lambda_*}}{\alpha} \delta\langle T_{00}^{\mathrm{CFT}}\rangle.
\label{eqn:curvature_eqn_nonlocal}
\end{equation}

All of our work thus far has been purely in the context of space, rather than spacetime.
We have not posited any form of Hamiltonian or time evolution.
Let us (somewhat optimistically) assume that the present framework can be adapted to a situation with conventional time evolution, and furthermore that the dynamics are such that an approximate notion of local Lorentz invariance holds. (Such an assumption is highly nontrivial; see \emph{e.g.} \cite{Collins:2004bp}.)
Thinking of our emergent space as some spatial slice of a Lorentzian spacetime manifold,
the spatial curvature can be related to the usual quantities in the Einstein tensor. In particular, for slices with vanishing extrinsic curvature we have
\be
 \mathcal{R}_p=2G_{00}(p).
\ee
Comparing to (\ref{eqn:curvaturemodularenergy}), we therefore find
\begin{equation}
G_{00} = \frac{2\pi\kappa_{\lambda_*}}{\alpha} \delta\langle T_{00}^{\mathrm{CFT}}\rangle.
\end{equation}
If we make the identification $2\pi\kappa_{\lambda_*}/\alpha\rightarrow 8\pi G$, this is nothing but the $00$-component of the semiclassical Einstein equation. The interaction strength is determined in part by the dimensionful constant $\alpha$ that relates entropy to area, similar to \cite{jacobson15}. If this reasoning is approximately true for all time-like observers traveling along $u^{\mu}$, then one can covariantize and arrive at the full equation 
\be
G_{\mu\nu} =  \frac{2\pi\kappa_{\lambda_*}}{\alpha}  \delta\langle T_{\mu\nu}^{\mathrm{CFT}}\rangle. 
\ee
This is the result for local perturbations using the smooth-manifold approach to the emergent geometry. An additional factor of $F'_{\Lambda}(0)/2$ would accompany nonlocal perturbations using the lossy channel on $A_p$, and an analogous equation can be derived from the Regge calculus approach using (\ref{eqn:RHrelation}).

An immediate consequence of our bulk emergent gravity program is that there is a bound on the change in entropy within a region, reminiscent of the Bekenstein and holographic bounds.
We have argued that positive energy corresponds to a decrease in the (full, UV) entropy, so we expect there to be an upper limit on the amount by which the entropy can decrease.
This is of course automatic, as the entropy is a nonnegative number.
Once the region $A_p$ is fixed, the maximum possible decrease in entropy that corresponds to positive ``mass-energy'' has to be bounded by the total entanglement entropy of $S_{A_p}$, which is proportional to the area $\area_{p}$ of the region.
More explicitly,
\begin{equation}
|\Delta S_{A_p}| \leq \alpha \area_{p}.
\end{equation}
This resembles the holographic entropy bound.
For an entropy change that saturates the bound, the vertices in regions $A_p$ and $\overline{A}_p$ become disconnected from the graph point of view. An embedding space for $\overline{A}_p$ that reflects this change now has a hole around region $A_p$.
Perturbations that increase the entropy are also bounded, but the bound scales with the volume of the region; we believe that configurations that saturate such a bound do not have a simple geometric interpretation.

\section{Discussion}\label{sec:conclusion}

We have examined how space can emerge from an abstract quantum state in Hilbert space, and how something like Einstein's equation (in the form of a relationship between curvature and energy) is a natural consequence of this bulk emergent gravity program.

We considered a particular family of quantum states, those that are ``redundancy-constrained'' in a given decomposition of Hilbert space.
For such states, a weighted graph that captures the entanglement structure can be constructed from the mutual information between different factors, and a manifold on which the graph can be (approximately) isometrically embedded is defined to be its emergent geometry. We presented specific implementations of the reconstruction framework using the classical multidimensional scaling algorithm for certain known area-law states. Both the dimension and the embedding coordinates for flat geometries can be found through the procedure.  At leading order, entanglement perturbations backreact on the emergent geometry, and allow modular energy to be associated with the spatial curvature. This relation is analogous to the semiclassical Einstein equation.

A crucial feature of this approach is that we work directly with quantum states, rather than by quantizing classical degrees of freedom. 
No semiclassical background or asymptotic boundary conditions are assumed, and the theory is manifestly finite (since regions of space are associated with finite-dimensional factors of Hilbert space). 
There is clearly a relation with approaches that derive geometry from the entanglement structure of a boundary dual theory, but the entanglement we examine is directly related to degrees of freedom in the emergent bulk spacetime.
Because lengths and other geometric quantities are determined by entanglement, a connection between perturbations of the quantum state and perturbations of the geometry appears automatically; in this sense, gravity appears to arise from quantum mechanics in a natural way.

Clearly, the framework is still very incomplete, and leaves much for future investigation. 
An important step in our procedure was assuming that we were given a preferred Hilbert-space decomposition $\mathcal{H} = \bigotimes \mathcal{H}_p$; ultimately we would like to be able to derive that decomposition rather than posit it.
Perhaps most importantly, our definition of distance in terms of mutual information is compatible with the behavior of field theories at low energies, but we would like to verify that this really is the ``distance'' we conventionally refer to in quantum field theory.
Ultimately that will require an investigation of the dynamics of these states.
An obvious next step is to define time evolution, either through the choice of an explicit Hamiltonian or by letting time itself emerge from the quantum state.
One important challenge will be to see whether approximately Lorentz-invariant dynamics can be recovered at low energies, and whether or not the finite nature of Hilbert space predicts testable deviations from exact Lorentz symmetry.
We might imagine that, given a state $|\psi\rangle$ whose geometry is constructed using entanglement, one can generate all time-slices using a known local Hamiltonian such that $|\psi\rangle$ is a low energy state. Alternatively, by working with mixed states one could adopt the thermal time hypothesis \cite{tth} and generate state-dependent time flow purely from the modular Hamiltonian, which is in principle attainable from just the density operator. 

To analyze the emergent geometries of states beyond redundancy-constraint, deeper understandings of the entropy data for subregions of different sizes will be important. One such case is manifest in the context of AdS/CFT correspondence, where entanglement entropy of different-sized balls in the CFT are needed to obtain bulk geometric information through a radon transform. One such approach may be to introduce additional structures on the graph and extend it to a tensor network.  The program of geometry from tensor networks has mostly been based on states with a high degree of symmetry, such that notions of length and curvature can be assigned through simple geodesic matching and tessellation of space. 
The results obtained here suggest that for tensor networks with small perturbations, one can also modify the geometric assignment accordingly, matching the change in correlation or entanglement to perturbation in geodesic lengths. A notion of (coarse) local curvature can also be defined on triangulated spaces using entanglement and Regge calculus,  which seems more natural for programs that relates network geometries to those of spacetime. 

The emergence of time evolution will also be  useful for the study of more complex behaviors related to entanglement perturbations. For instance, one can examine the interactions among multiple perturbations created in some local region. If the model is truly gravitational, the time evolution experiment should be consistent with our knowledge of gravitational dynamics. 
It will also be interesting to study the redundancy-constrained deformations of states beyond perturbative limit. Intuitively, we expect the emergence of a classical wormhole geometry by nonlocally entangling large number of degrees of freedom in a coherent manner. 
One can also examine purely quantum phenomena outside the context of classical Einstein gravity, including black-hole entropy and evaporation, using mutual information rather than classical spacetime geometry.

\section*{Acknowledgements}

We would like to thank Ning Bao, Aidan Chatwin-Davies, Bartek Czech,  Nick Hunter-Jones, Shaun Maguire, Hirosi Ooguri, John Preskill, Jason Pollack, and Brian Swingle for helpful discussions. C.C. would like to thank Ning Bao copiously for his suggestions and support throughout the course of this project. C.C. also thanks the orgranizers of the YITP long term workshop on ``Quantum Information in String Theory and Many-body Systems''. 
This research is funded in part by the Walter Burke Institute for Theoretical Physics at Caltech, by DOE grant DE-SC0011632, by the Foundational Questions Institute, by the Gordon and Betty Moore Foundation through Grant 776 to the Caltech Moore Center for Theoretical Cosmology and Physics, and by the John Simon Guggenheim Memorial Foundation. S.M. acknowledges funding provided by the Institute for Quantum Information and Matter, an NSF Physics Frontiers Center (NSF Grant PHY-1125565) with support of the Gordon and Betty Moore Foundation (GBMF-2644).

\appendix

\section{Redundancy-Constraint and Coarse-graining} \label{app:coarsegraining}

In this appendix we consider how to construct a coarse-grained decomposition of Hilbert space that is redundancy-constrained (RC), as defined in Section~\ref{sec:RC}, when an initial fine-grained one is not.

Given some state $|\psi_0\rangle\in\mathcal{H}$ and some fixed Hilbert space decomposition $\mathcal{H}=\bigotimes_{i}^M\mathcal{H}_i$ for $M$ sufficiently large, we first create a network represented by a graph $G_0=(V_0,E_0)$. The graph has $N$ vertices labelled by $i$, and each edge $\{i,j\}$ is weighted by $I(i\co j)$, where $I(i\co j)$ is the mutual information of partitions $i$ and $j$. If the resulting graph is RC to the desired degree of accuracy, no further coarse-graining is needed. 

If not, consider the set of all partitioning schemes $\mathcal{C}$ for a coarse-grained decomposition of the Hilbert space such that $\mathcal{H}=\bigotimes_{p}^{N}\mathcal{H}_p$. For each scheme $\mathcal{S}\in\mathcal{C}$, we require that $N\leq M$ for some sufficiently large $N$ so that non-trivial entanglement structure is still allowed. 
Each partition $\mathcal{S}=\{\{i_1,i_2,\dots\},\{i_k,i_{k+1},\dots\},\dots\}$ corresponds to constructing a more coarse-grained decomposition of the Hilbert space by taking the union of original subfactors; that is, for each $s_p\in\mathcal{S}$, $\mathcal{H}_p=\bigotimes_{i\in s_p}\mathcal{H}_i$. 

A partition is RC-valid if the mutual-information-weighted-network $G=(V,E)$ based on the coarse-grained decomposition is redundancy-constrained. While there is no obvious way to choose the best coarse-graining scheme at this point, it is natural to consider the most uniform partitioning, so that all Hilbert-space subfactors have approximately equal dimensions. 

If no such coarse-graining can be found for $N$ reasonably large, then the procedure fails and we are forced to conclude that the given state cannot be cast in RC form in a simple way. We do not claim, however, that it doesn't admit a simple geometric description, as this is clearly false from our knowledge of AdS/CFT. Reconstructing the geometry from such states are interesting problems. 

A more specific method for coarse-graining can also be constructed using network renormalization. While the algorithm is less computationally intensive, it fails for certain states if the original decomposition yields little useful information. For instance, it fails for the ground state of Toric code on a square lattice where the given Hilbert space decomposition is the usual tensor product of spin-1/2 degree of freedom on each link. It is, however, useful for certain finitely correlated states and/or those of typical condensed matter system at scales larger than the correlation length. It also works for the toric code if the decomposition is more coarse-grained.

For such states and decompositions, we follow network renormalization procedure by again constructing the mutual-information-weighted-network $G_0=(V_0,E_0)$. Assume $G_0$ is connected; in the case when $G_0$ has multiple large disconnected components, one can simply perform the procedure separately for each connected component.

Then proceed to define a metric $\tilde{d}_0(i,j)$ on $G_0$ in the same manner as (\ref{eqn:metric}), with $\ell_{\mathrm{RC}}\rightarrow \ell_0$. 
Then, for any vertex $v$ on the graph, we consider an $\epsilon$-ball $B_{\epsilon}(v)$ such that
\begin{equation}
B_{\epsilon}(v) = \{ v'\in V | \tilde{d}(v,v')\leq \epsilon\},
\end{equation}
where $\tilde{d}(v,v')$ is a metric defined on the set of vertices $V$. Seed the entire graph with points like $v$ until the whole graph is covered by $\epsilon$-balls. Choose a minimum cover and compute the entanglement entropy for each the union of subregions in each ball to generate a coarse-grained graph $G_1$, where each vertex now is labeled by the union of the original subregions and the edge weights are given by the mutual informations of the coarse-grained subregions. Repeating the coarse-graining procedure until all higher subregions $B_X$ can be well approximated by the cut function (\ref{eqn:cutfcn}), we then label the coarse-grained subregion at this scale to be $A_p$ for $p\in \mathcal{S}$, and label the corresponding coarse-grained network with $A_p$ also. The resulting state will be, to a good approximation, redundancy-constrained.

\section{Entanglement Perturbations and Coarse Curvature}
\label{app:coarse}

In this appendix we consider an alternative approach to calculating the curvature induced by an entanglement perturbation, working directly with the discrete graph rather than finding an embedding Riemannian manifold. 
Here we use the techniques in Regge calculus \cite{regge}, in which the sense of spatial curvature is determined by deficit angles.  

For a space of fixed integral Hausdorff dimension $D$ obtained from (\ref{eqn:HausdorffDim}), consider a vertex $p$ and construct a local triangulation, if one exists. We say the space is $r$-locally triangulable at $p$ if one can construct a abstract simplicial complex $\mathcal{K}$, where the simplices are sets of vertices in the metric subspace near $p$, by imposing a distance cutoff $r$, and if there exists an isometry (with respect to the metric distance) that maps $\mathcal{K}$ to a geometric simplicial complex $K$ where inter-vertex distances of the metric subspace are preserved. If $K$ is also a simplicial manifold we can proceed to define Regge curvature.

Select a codimension-2 simplex $X\ni p$ as a hinge, its volume given by (\ref{eqn:simplicialvolume}).  As the simplices in $K$ are equipped with the usual Euclidean inner product, angles can be defined and deficit angle at the hinge $\delta(X)=2\pi-\theta(X)$ can computed using the inner product structure in Euclidean space. Here we define 
  \begin{equation}
   \theta(X) = \sum_i \phi_i(X),
 \end{equation}
where $\phi_i(X)$ is the angle between the unique two faces of a simplex containing the hinge $X$.
In the case of a $D$-dimensional area-law system where $I(p\co q)\neq 0$, construct a simplex by considering the $n$ shortest distances. 
The curvature is then related to the deficit angles $\delta_i$ in the region by
  \begin{equation}
   R_T = \sum_i\delta_i L_i,
  \label{eqn:coarseR}
  \end{equation}
where $L_i$ are the volumes of the codimension-2 hinges at which the curvature is concentrated. For $D=2$, the hinge is a point, and the curvature is given by the deficit angle, where we set $L_i=1$. In the continuum limit, (\ref{eqn:coarseR}) becomes $\int d^D x \sqrt{g} \mathcal{R}$, where $\mathcal{R}$ is the scalar curvature. In the case of emergent Euclidean (flat) space, we require $R_T=0$.

Let us consider the effect of an entanglement perturbation on the geometry of a distance-function graph $\widetilde G$, using this technique.
Again, the forms for both local and nonlocal perturbations are similar up to an overall factor. 

Since the original deficit angle $\delta_p=0$ in flat space,  for $p$ lying on the hinge, we may consider the angular deficit produced by varying each elongated edge connected to $p$. For each such simplex $S$ connected to the hinge, the deficit at $p$ induced by varying the length $l_j$ of each edge in the simplex assumes, at leading order, the form
\begin{equation}
\delta_p^{(S)} = \frac{\delta l_j} {\ell^{(S)}_j(l_1,l_2,\dots)}+O((\delta l_j/l_j^{(S)})^2),
\end{equation}
where $l_j$ denotes the $j$th edge length of the $D$-simplex $S$. $\ell^{(S)}_j(l_1,l_2,\dots)$ is a function that has dimension of length and depends on the edge that is varied, as well as all the edge lengths that connect the vertices of $S$. 

The overall deficit in a triangulation where all edges have roughly the same (unperturbed) length $\ell$ is 
\begin{align}
\Delta_p &= \sum_{j,S} \frac{\delta l_j} {\ell^{(S)}_j}+O((\delta l_j/l_j^{(S)})^2)\\
	&= \mathcal{N}_p(D) \frac{\delta \ell }{\ell}+O((\delta \ell/\ell)^2),
\end{align}
where $\mathcal{N}_p(D)$ depends on the simplices in the triangulation. For example, if equilateral triangles with sides $\ell$ are used to triangulate the 2-dimensional flat space around $p$, then $\mathcal{N}_p (D=2) = 12/\sqrt{3}$ in the case where all edges emanating from $p$ vary by the same amount $\delta \ell$ under the entanglement perturbation.
 Note that in dimension greater than 2, there is no uniform tiling such that all edges are equal, hence the approximation in some average sense.

To estimate the coarse curvature at $p$, take $L_p \sim \ell^{D-2}$ as the volume of the codimension-2 hinge. $\delta \ell$ is identified with $\delta \tilde{d}(p,q)$ for the change in distance between adjacent vertices, and $\ell \approx \tilde{d}(p,q)$ for all $q$ immediately adjacent to $p$ in the triangulation. The total coarse curvature is given by
\begin{equation}
R_c=\Delta_p L_p\sim \mathcal{N}_p(D)\ell^{D-3}  \delta \ell.
\end{equation}
On the other hand, we know from  \cite{regge} that in the continuum limit, if the metric $g$ is approximately constant in the small region, $R_c \rightarrow \ell^D \mathcal{R}$, where $\mathcal{R}$ is the average coarse scalar curvature of the space contained in the small region with approximate volume $\ell^D$.  Applying (\ref{eqn:distancemutualinfo}) and the EFL, the average coarse curvature in the region is
\begin{equation}
\mathcal{R}=Z\delta\langle H\rangle.
\label{eqn:RHrelation}
\end{equation}
Here, for a perturbation induced by a unitary $U_{A_p\overline{A}_p}$, the constant $Z=\ell_{\mathrm{RC}}^{-2}\xi(p,q,D)$ depends on the triangulation, strength of entanglement, and choice of coarse-graining. The factor $\xi(p,q,D)\propto |\mathcal{N}_p(D)\Phi'(i(p\co q))|/I_0(p\co q)$, which parametrizes all the order-one constants that enter into the process of averaging, can be explicitly computed once the triangulation and the inverse function $\Phi$ are known. We have taken $\ell = \gamma \ell_{\mathrm{RC}}$ to denote the average radius of the region, as before. Note that this is consistent with (\ref{eqn:curvatureLengthExcess}) up to the dimension- and triangulation-dependent factors. 

For a nonlocal perturbation through a channel $\Lambda$, because the unitary only acts on $A_p$ and the ancilla, the mutual information $\delta i(s\co q)=0$, and hence the distance functions $\tilde{d}(s,q)$ are invariant for all $s, q\neq p$. Only legs emanating from the vertex $p$ in the triangulation are varied. The unitary perturbing map has a wider range of possible consequences, as its form is unspecified. Although the values of $Z$ may be different depending on the specific map used, the formalism remain the same. As a result, for nonlocal perturbations with channel $\Lambda$, we have an equation of the same form, with $Z\rightarrow ZF'_{\Lambda}(0)/2$. In particular, if we restrict $U_{A_p\overline{A}_p}$ to only remove entanglement symmetrically near the boundary of the region $A_p$ as before, the values of $Z$ will be the same.


\bibliography{emergentspace}
\bibliographystyle{JHEP}


\end{document}